\documentclass[trackchanges]{aastex62}
\usepackage{amssymb}
\usepackage{array}
\usepackage{graphicx}
\usepackage{multirow}

\newcommand\MyBox[1]{
  \fbox{\lower0.375cm
    \vbox to 0.85cm{\vfil
      \hbox to 0.85cm{\hfil\parbox{0.85cm}{\centering #1}\hfil}
      \vfil}
  }
}

\begin{document}

\title{A Random Forest Approach to Identifying Young Stellar Object Candidates in the Lupus Star-Forming Region}
\author {Elizabeth Melton}
\affil{Department of Astronomy and Astrophysics, The Pennsylvania State University, University Park, PA 16802, USA}

\begin{abstract}
The identification and characterization of stellar members within a star-forming region are critical to many aspects of star formation, including formalization of the initial mass function, circumstellar disk evolution and star-formation history.  Previous surveys of the Lupus star-forming region have identified members through infrared excess and accretion signatures.  We use machine learning to identify new candidate members of Lupus based on surveys from two space-based observatories: ESA's \emph{Gaia} and NASA's \emph{Spitzer}.  
Astrometric measurements from \emph{Gaia}'s Data Release 2 and astrometric and photometric data from the Infrared Array Camera (IRAC) on the \emph{Spitzer} Space Telescope, as well as from other surveys, are compiled into a catalog for the Random Forest (RF) classifier.  
The RF classifiers are tested to find the best features, membership list, non-membership identification scheme, imputation method, training set class weighting and method of dealing with class imbalance within the data.  
We list 27 candidate members of the Lupus star-forming region for spectroscopic follow-up.  Most of the candidates lie in Clouds V and VI, where only one confirmed member of Lupus  was previously known. These clouds likely represent a slightly older population of star-formation.
\end{abstract}


\section{Introduction \label{sec:intro}}

The stellar populations associated with star-forming regions are used to constrain stellar evolution theories, examine the progression of star-formation, and characterize protoplanetary disks.  
In order to get a clear picture, it is important to have as complete a sample as possible of all of the members of a star-forming region.  
A wide variety of methods used for identifying young stars in star-forming regions: emission from active young stars \citep{krautter97}, infrared excess from circumstellar disks \citep{allers06, merin08, comeron09} and similar kinematic motion \citep{lopezMarti05}.  
As the use of wide-field and all-sky surveys (\emph{Gaia}, TESS, LSST, WFIRST) covering hundreds of millions and billions of sources grows, so does the ability to combine information as well as examine a large number of sources.  

Machine learning techniques can be and have been used to identify candidate young stellar objects.  Both \citet{miettinen18} and \citet{marton19} explored a variety of classification techniques to attempt and find the best technique for identifying young stellar objects in large data sets.  In both studies, the Random Forest (RF) classifier technique performed very well compared with the other options (neural networks, support vector machines, na\"{i}ve bayes, k-nearest neighbor, etc.); the RF performed second only to gradient boosting in \citet{miettinen18} and best in \citet{marton19}.   However, in these studies, young stellar objects were classified solely on photometric indices.  Here we use both astrometric and photometric data to identify candidate members of a star-forming region.  We use RF for this purpose because it easily handles data of differing units without running into the problems present in some other types of machine learning classifiers. 

We use RF to identify new candidates of the Lupus star-forming region after comparing a number of construction methods. Lupus is part of the Gould Belt, the nearby ring of star-forming regions surrounding the solar neighborhood.  Lupus lies near the Scorpius-Centaurus OB Association, between Upper Scorpius and Upper Centaurus-Lupus \citep{comeron08}.   
It is made up of multiple clouds with various levels of star formation activity seen within them.  
Lupus III contains one of the richest and densest clusters of T Tauri stars known, while only sparse star formation is observed in Lupus I. 
Lupus clouds V and VI have practically no observed star formation.  
Lupus was included in the \emph{Spitzer} ``From Molecular Cores to Planet-forming Disks'' (c2d) legacy program to study the lifetimes and distribution of circumstellar disks and protostar cores in star-forming regions \citep{evans03, evans09}.  The \emph{Herschel Space Observatory} has also been used to study pre-stellar and stellar cores in the Lupus I, III and IV clouds \citep{rygl13, benedettini18} as well the characterization of disks \citep{bustamante15}.
Lupus has a slight overabundance of low-mass stars as compared with other nearby star-forming regions \citep{comeron08, merin08, mortier11, alcala17}. 

The search for low-mass stars and brown dwarfs in Lupus has been conducted primarily through photometric surveys for the infrared excess from circumstellar disks \citep{allers06, chapman07, merin08, spezzi11, dunham15}.  The main clouds are well covered by the Infrared Array Camera (IRAC) on the \textit{Spitzer Space Telescope} \citep{fazio04}.  \citet{esplin16} developed astrometric distortion corrections that allow the use of multi-epoch IRAC images to measure precise proper motions.  
These have been successfully used to measure proper motions for the Taurus and Chameleon I star-forming regions respectively \citep{esplin17_taurus, esplin17_cham} to identify previously unknown low-mass and even brown dwarf (M $\lesssim$ 0.075 $M_{\sun}$) members of these regions.  
Here, the same proper motion method is applied to the Lupus star-forming region to help identify candidate members for spectroscopic follow-up.  We also include proper motion and parallax from the \emph{Gaia} second data release.  Kinematic studies have been conducted for the Lupus star-forming region before \citep{lopezMarti11, galli13}, but they did not include photometric data in their assessments along with the astrometric data.

Section \ref{sec:data} covers the compilation of the data catalog, including our cleaning restrictions we apply to the data.  
Section \ref{sec:rand_forest} describes RF classifiers and the different input parameters over which the RF machine learning classifiers are tested.  
Section \ref{sec:rand_forest_comp} details the performance comparison of different classifiers, the best of which identifies candidate members of the Lupus star-forming region in Section \ref{sec:cands}.   In Section \ref{sec:ext} we use the best RF tuning parameters to extend the search for members of the Lupus star-forming region to include the area surrounding the main clouds. Finally, Section \ref{sec:discuss} explores the qualities of the candidate members.

\section{Classification Input Catalogs \label{sec:data}}

A source list of objects toward the Lupus star-forming region is constructed from archived IRAC images in the \emph{Spitzer} Heritage Archive.  
These images span a range of over 10 yr and cover five distinct regions of the Lupus star-forming region: providing coverage over Lupus Clouds I, III-VI. 

\begin{deluxetable}{ccc}
\tablecaption{Approximate boundaries of the IRAC observations of Lupus Clouds. \label{tab:cloud_boundary}}
\tablehead{\colhead{Cloud} & \colhead{RA range} & \colhead{DEC range}
}
\startdata
I & 15$^{h}$35$^{m}$ - 15$^{h}$48$^{m}$ & -33$\degr$ - -35$\degr$ 
\\
III & 16$^{h}$07$^{m}$ - 16$^{h}$16$^{m}$ & -37$\degr$50$\arcmin$ to -39$\degr$40$\arcmin$ \\
IV & 15$^{h}$58$^{m}$ - 16$^{h}$05$^{m}$ & -41$\degr$30$\arcmin$ to -42$\degr$10$\arcmin$\\
V & 16$^{h}$17$^{m}$ - 16$^{h}$26$^{m}$ & -36$\degr$30$\arcmin$ to -38$\degr$10$\arcmin$ \\
VI & 16$^{h}$19$^{m}$ - 16$^{h}$30$^{m}$ & -39$\degr$ to -42$\degr$ \\
\enddata
\end{deluxetable}

\subsection{Spitzer IRAC Astrometry and Photometry\label{sec:irac}}

Between 2004 and 2014, the IRAC instrument on the \emph{Spitzer Space Telescope} to image the Lupus star-forming region (see Table \ref{tab:epochs}). 
Forty-four observing runs were performed in 2004 (mostly as part of the c2d legacy program; \citealt{evans03, evans09}), one observing run in 2006, 12 were conducted in 2007, four observing runs in 2008 and two observing runs in 2011.  
The final mosaic images from these early observing runs did not have much overlap between neighboring images within each observing run.  
In contrast, the 28 observing runs performed in 2013 and six observing runs from 2014 by Kraus, A. had large overlaps between images.  These observing runs are ideal for measuring astrometry due to the large overlap between neighboring images.

\begin{deluxetable}{lllc}
\tablecaption{IRAC Epochs of Lupus \label{tab:epochs}}
\tablehead{
\colhead{AOR\tablenotemark{a}} & \colhead{PID\tablenotemark{b}} & \colhead{PI} & \colhead{Epoch} 
}
\startdata
3652608 & 6 & G. Fazio & 2004.151 \\
4921088 & 94 & C. Lawrence & 2004.151 \\
4921344 & 94 & C. Lawrence & 2004.151 \\
5669376 & 173 & N. Evans & 2004.675 \\
5670656 & 173 & N. Evans & 2004.675 \\
\enddata
\tablenotetext{a}{Astronomical Observing Request ID}
\tablenotetext{b}{Program ID}
\tablecomments{Table \ref{tab:epochs} is published in its entirety in the machine-readable format. A portion is shown here for guidance regarding its form and content.}
\end{deluxetable}

The archived images span IRAC's cryogenic and post-cryogenic cycles.  IRAC's cryogenic cycle, running from immediately post-launch until May 2009, contained broadband filters at 3.6, 4.5, 5.8 and 8.0 $\micron$, hereafter written as [3.6], [4.5], [5.8] and [8.0] \citep{fazio04}. 
The mean full-width half-maximum (FWHM) of point sources for [3.6]-[8.0] ranges from  $1\farcs66$ to $1\farcs98$ as wavelength increases \citep{fazio04}.  
After the depletion of the liquid helium aboard the \emph{Spitzer Space Telescope} and the start of the post-cryogenic cycles, the instrument operated in only the [3.6] and [4.5] bands.  
In order to retain the longer baseline for the proper motion measurements and take advantage of the smaller FWHM for point sources, only the [3.6] and [4.5] bands are utilized here.  

\begin{figure*}
\epsscale{0.8}
\plotone{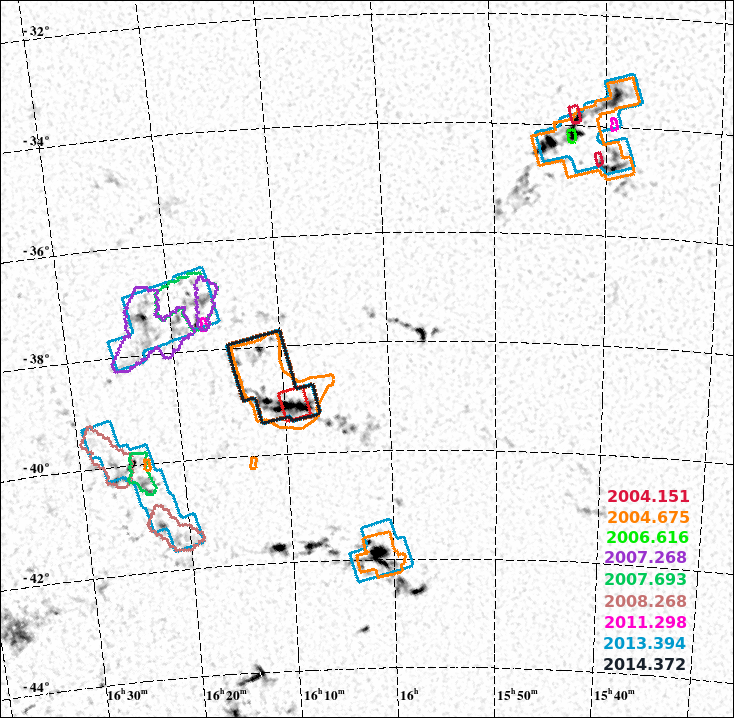}
\caption{Spatial coverage of each of the epochs of IRAC image data over the Lupus region.  Five of the clouds within the Lupus star-forming region (see Table \ref{tab:cloud_boundary}) have multi-epoch coverage; each of the regions are covered by epoch 2013.394 or epoch 2014.372. The epoch coverage is overlaid on an extinction map of the Lupus star-forming region by \cite{dobashi05}. \label{fig:epoch_boundaries}} 
\end{figure*}

The Astronomical Observing Requisition IDs (AORs), Program IDs (PIDs), principal investigators and mean observation date for each epoch are listed in Table \ref{tab:epochs}.  
Figure \ref{fig:epoch_boundaries} shows the coverage provided by each epoch overlaid on an extinction map of the Lupus star-forming region \citet{dobashi05}.  
The \emph{Spitzer} IRAC coverage of Lupus covers five of the main clouds: Lupus I, III-VI.  The  approximate boundary of each cloud is given in Table \ref{tab:cloud_boundary} \citep{comeron08}.  

For each image, the point-response-function (PRF) fitting routine in the Astronomical Point source EXtractor (APEX, \citealt{makovoz05}) extracts sources.   
Sources that are nearly saturated in the image, sources with $F_{\nu}$/(exposure time) $>$ 0.728 for [3.6] images or $F_{\nu}$/(exposure time) $>$ 0.822 for [4.5] images, are removed due to the inability to precisely measure their astrometry.  
A master list of sources for the epoch is created by running the PRF fitting routine of APEX on a master mosaic created from a mosaic of all the images within each epoch and  corrected for distortions in the position of the sources.  
The time-dependent distortion in the plate scale of IRAC is corrected for using the same interpolation method presented in \citet{esplin16} for each epoch.  
Photometry in the [3.6] and [4.5] passband filters is calculated from source flux measurements using the parameters from the IRAC instrument handbook \citep{irac15}.

The procedure developed in \citet{esplin16} and demonstrated in \citet{esplin17_taurus} and \citet{esplin17_cham} is utilized to measure proper motion for the sources.  
\citet{esplin16} demonstrated that astrometry as precise as 70 mas for sources with a low signal-to-noise ratio (S/N) up to 20 mas for sources with a high S/N is possible for the [3.6] and [4.5] IRAC bands.  

\begin{figure*}
\epsscale{1.15}
\plotone{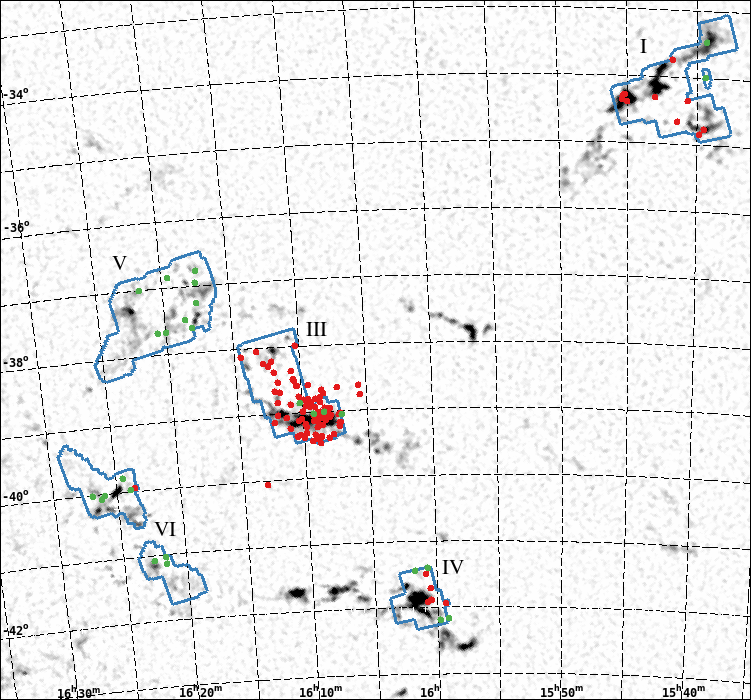}
\caption{ Positions of the members of the Lupus star-forming region (red dots) and the proposed candidate members (green dots) overlaid on an extinction map of the Lupus star-forming region \citep{dobashi05}.  The area of the Lupus star-forming region with good IRAC proper motion is outlined in blue.   \label{fig:pm_final}}
\end{figure*}

In order to minimize the number of spurious measurements for sources, we set limits on the errors of the proper motion and S/N following the same procedure described in \citet{esplin17_taurus}.  Of the 716,530 sources within the IRAC field, only 85,431 have good proper motion measurements (i.e. have sufficient number of detections over multiple epochs, are unsaturated and meet our S/N and error limits).  
Figure \ref{fig:pm_final} shows the coverage of the final IRAC proper motion catalog within the five cloud regions.  

\subsection{Gaia Astrometry\label{sec:gaia}}

IRAC is well suited to detecting low-mass sources and sources in star-forming regions because of the low extinction in the infrared regime and the cool nature of low-mass sources and substellar objects.  
We include \emph{Gaia} data in our analysis to further refine candidate selection with its more precise proper motion measurements and parallax.  

The \emph{Gaia} mission second data release contains proper motion measurements of sources, parallax measurements and optical broadband photometry ($G$), covering 0.330 to 1.050 $\micron$ \citep{gaia18, lindegren18}.     
Overall \emph{Gaia} is less sensitive to low-mass sources and those that have been been  reddened by high extinction but its proper motion and parallax measurements complement the IRAC proper motions measured in Section \ref{sec:irac} and allow for more thorough constraints on members and candidate members.  

The list of sources from the on-cloud catalog is cross-matched with the second data release of \emph{Gaia} with a radius of $1\farcs5$ on the sky.   
In order to identify sources for which a good \emph{Gaia} astrometric solution was found, the astrometric excess noise, the difference between the observations of the source and the astrometric solution, and the astrometric goodness of fit parameter, the Gaussianized $\chi$-squared value of the astrometric solution \citep{lindegren12} are examined.  
Average shifted histograms are used to create a two-dimensional smoothed density estimation of the astrometric excess noise as a function of the astrometric goodness of fit using the CRAN package ``ash" \citep{r13, scott15}.  
The \emph{Gaia} astrometry for sources that fall outside of the contours for a local logarithmic surface density of -4.25 are not included in the final catalog covering the five cloud regions.  

Parallax measurements for sources that have a negative parallax are removed from the catalog.  This will not bias our results because we are looking for members of the Lupus star-forming region, which should have a parallax between 5 and 6.667 mas \citep{comeron08}.    
Parallax measurements with errors greater than 10\% of the parallax measurement are also removed.  These measures in our input catalog effectively guard against two of the large problems pointed out in \citet{bailer-jones18} in translating the \emph{Gaia} parallax measurements to distance measurements; the positivity constraint of distance and the low S/N of the parallaxes.  The parallax measurements that we are interested in are in a small range and so these measures will not affect the final results.

\subsection{2MASS and DENIS Photometry\label{sec:phot}}

$J$ (1.25 $\micron$), $H$ (1.65 $\micron$) and $K_{s}$ (2.15 $\micron$) band photometry from the Two Micron All Sky Survey \citep[2MASS, ][]{skrutskie06} and $i$ (0.82 $\micron$), $J$ (1.25 $\micron$) and $K_{s}$ (2.15 $\micron$) band photometry from the Deep Near-Infrared Survey of the southern sky \citep[DENIS, ][]{epchtein99} are incorporated by matching the source list to the photometric catalogs.  
The $J$ filters from 2MASS and DENIS are treated as the same bandpass and combined, as are the $K_{s}$ filters.  
We remove 2MASS and DENIS photometry data from the on-cloud catalog with errors above 0.1 magnitude.    

Extinctions for individual sources are estimated by dereddening the $J$ - $H$ color for each source as a function of $H$ to the typical locus of young stars at the distance of Lupus.  
The distance to Lupus \citep{comeron08} is very similar to the Chameleon I star-forming region, therefore we adopt the same extinction estimation as \citet{esplin17_cham}  adopted.  
The reddening relations from \citet{schlafly16} are used to correct the photometry for extinction in the \emph{Gaia} $G$ band through the IRAC [4.5] band along with the $A_{H}$/$A_{K_{s}}$ = 1.55 relation from \citet{indebetouw05}.  
No extinction correction is applied to sources lacking either $J$ or $H$ photometry.

\subsection{Final On-Cloud Catalog}

In total, the catalog within the five cloud regions has 716,530 sources (source list drawn from IRAC images) and 14 features: position, parallax, IRAC proper motion, \emph{Gaia} proper motion, $G$, $i$, $J$, $H$, $K_{s}$, [3.6] and [4.5] band photometry.  
Most sources have at least one feature measurement missing for a variety of reasons, including but not limited to instrument sensitivity, instrument coverage and exclusion due to excess error.  
A total of 14,795 sources have parallax measurements, 85,431 sources have IRAC proper motion, 393,357 sources have \emph{Gaia} proper motion, 463,838 sources have $G$ photometry, 91,312 sources have \emph{i} photometry, 164,429 sources have $J$ photometry, 152,406 sources have $H$ photometry, 119,813 sources have $K_{s}$ photometry, 485,909 have [3.6] photometry and 489,156 sources have [4.5] photometry.  

\subsection{Extended Region Catalog}

We create an extended catalog of sources to search for candidate members of the Lupus star-forming region that may have dynamically ejected at birth and traveled away from the main area of the clouds or represent an older population of star-formation that may have become more dispersed \citep{feigelson96, comeron13}.  The extended catalog consists only of data from the \emph{Gaia} second data release, 2MASS and DENIS.  All data were cleaned and combined into the extended catalog as described above for the on-cloud catalog.  Table \ref{tab:ext_boundary} gives the boundaries of the extended catalog by cloud.  The extended catalog is constructed so that each cloud is covered by at least 0.5$\degr$ beyond the IRAC coverage boundaries.  Sources from the on-cloud catalog are removed from the extended catalog. 

\begin{deluxetable}{ccc}
\tablecaption{Boundaries of the extended catalog around each of the Lupus Clouds. \label{tab:ext_boundary}}
\tablehead{\colhead{Cloud} & \colhead{RA range} & \colhead{DEC range}
}
\startdata
I & 15$^{h}$30$^{m}$ - 15$^{h}$55$^{m}$ & -36$\degr$ - -32$\degr$ 
\\
III & 16$^{h}$00$^{m}$ - 16$^{h}$20$^{m}$ & -41$\degr$ - -36$\degr$ \\
IV & 15$^{h}$55$^{m}$ - 16$^{h}$10$^{m}$ & -43$\degr$ - -40$\degr$\\
V & 16$^{h}$10$^{m}$ - 16$^{h}$35$^{m}$ & -39$\degr$ - -35$\degr$ \\
VI & 16$^{h}$15$^{m}$ - 16$^{h}$35$^{m}$ & -43$\degr$ - -38$\degr$ \\
\enddata
\end{deluxetable}

The extended catalog mainly contains off-cloud sources, but does contain on-cloud sources as well.  The on-cloud catalog includes only those sources that have good IRAC flux measurements in either the [3.6] or [4.5] band and do not include sources that may have been contaminated, over-exposed, etc.  In total, the extended catalog is made up of 7,931,039 sources (over 10 times the number of sources in the on-cloud catalog) and 10 features (seven with error measurements): position (right ascension and declination), parallax, \emph{Gaia} proper motion (right ascension and declination), $G$, $i$, $J$, $H$, and $K_{s}$ band photometry.  Of these sources, 139,367 sources have parallax measurements, 5,503,486 sources have \emph{Gaia} proper motion, 7,608,531 sources have $G$ photometry, 764,194 sources have \emph{i} photometry, 1,364,381 sources have $J$ photometry, 1,157,110 sources have $H$ photometry, and 851,533 sources have $K_{s}$ photometry.

\section{Random Forest\label{sec:rand_forest}}

Developed by \citet{breiman01}, the RF is a classifier which uses the majority class predicted by a collection of decision trees to assign a class label.  
Each decision tree is grown independently from the others and at each node of the tree, only a randomly selected subset of data features are used to split the data.  
This ensures that each decision tree in the RF classifier is unique.  
The RF is a very robust classifier, shown to be able to obtain high accuracy with minimal tuning of the internal classifier parameters.  
RFs are well suited to astronomical classification because they combine data of different units and do not suffer from scaling issues with the input parameters. 

The CRAN ``randomForestSRC" package \citep{ishwaran18} in the R statistical software environment \citep{r13} implements our RF.  
Of the variety of RF packages to choose from, we choose this one for its imputation procedures \citep{tang17} handling missing data. 
The implementation of RFs within the package, designed for the analysis of survival data in a bio-statistical problem, is consistent with original prescriptions of the RF process \citep{ishwaran08}.   

RF classifiers have parameters that can be tuned for optimal performance.  
This study focuses on three areas of possible tuning parameters: the training data set labels, the catalog features passed to the RF and the imputation method incorporated.  

\subsection{Training Set Options\label{sec:training_set}}

While the RF is fairly robust to noise in the labels of the training set \citep{breiman01}, the properties of the sources labeled as members for training should be such that it separates the properties of the members from the larger population of background stars.  
The choice of labeled members, labeled non-members and the relative abundance of each in the training set affect the effectiveness of the RF.

\subsubsection{Member List\label{sec:mem}}

A literature search performed for stars that were spectroscopically confirmed as members of Lupus finds 128 sources that fall within the IRAC image coverage.  
Table \ref{tab:members} lists the previously identified members along with their spectral types and the appropriate references. 
The catalog data (either on-cloud or extended as appropriate) for the previously identified members are presented in Table \ref{tab:members_info}. 

\begin{deluxetable*}{llcclh}
\tabletypesize{\scriptsize}
\tablecaption{Literature search results for previously identified spectroscopically confirmed members of the Lupus star-forming region with their spectral types and reference(s).  \label{tab:members}}
\tablehead{
\colhead{} & \colhead{Name} & \colhead{Spec.}  & \colhead{Ref.\tablenotemark{a}} & \colhead{Lada \tablenotemark{a}} & \nocolhead{Mass\tablenotemark{a}}\\
 \colhead{} & \colhead{} & \colhead{Type} & \colhead{SpTyp} & \colhead{Class} & \nocolhead{}
}
\startdata
1 & Sz 65 & M0/K7 & 1/11 & Class II (5, 11) & 0.46 (1) \\
2 & Sz 66 & M3 & 1, 8, 11 & Class II (5, 10, 11) & 0.37 (1), 0.45 (8), 0.31 (11) \\
3 & 2MASS J15394637-3451027 & K4 & 2 & \nodata   &  \nodata  \\
4 & Sz 67 & M4 & 1 & Class III (5) & 0.15 (1) \\
5 & 2MASS J15412189-3440150 & M2.5 & 2 & No IR excess (12) & \nodata   \\
\enddata
\tablenotetext{a}{Reference(s): 1. \citet{hughes94}, 2. \citet{krautter97}, 3. \citet{comeron03}, 4. \citet{allen07}, 5. \citet{merin08}, 6. \citet{mortier11}, 7. \citet{comeron13}, 8. \citet{alcala14}, 9. \citet{muzic14}, 10. \citet{dunham15}, 11. \citet{alcala17}, 12. this work}
\tablecomments{ Previously identified members in the extended catalog are listed at the end and have an I.D. beginning with "E-". }
\tablecomments{Table \ref{tab:members} is published in its entirety in the machine-readable format. A portion is shown here for guidance regarding its form and content.}
\end{deluxetable*}

\begin{splitdeluxetable*}{lccrrrrrBrrrrrrrcl}
\tablecaption{Catalog information for the previously identified members in Table \ref{tab:members}. \label{tab:members_info}}
\tablehead{
\colhead{} & \colhead{RA} & \colhead{DEC} & \colhead{$\mu_{\alpha}$ (IRAC)} &  \colhead{$\mu_{\delta}$ (IRAC)} & \colhead{$\mu_{\alpha}$ (\emph{Gaia})} &  \colhead{$\mu_{\delta}$ (\emph{Gaia})}  & \colhead{Parallax} & \colhead{G} & \colhead{\emph{i}} & \colhead{J} & \colhead{H} & \colhead{K} & \colhead{[3.6]} & \colhead{[4.5]} &\colhead{P$_{RF}$} & \colhead{}\\
\colhead{} & \colhead{hh:mm:ss.ss} & \colhead{dd:mm:ss.s} & \colhead{mas yr$^{-1}$} & \colhead{mas yr$^{-1}$} & \colhead{mas yr$^{-1}$} & \colhead{mas yr$^{-1}$} & \colhead{mas} & \colhead{mag.}  & \colhead{mag.} & \colhead{mag.} & \colhead{mag.} & \colhead{mag.} & \colhead{mag.} & \colhead{mag.} &\colhead{} & \colhead{}
}
\startdata
1 & 15:39:27.77 & -34:46:17.3 &     \nodata         &     \nodata          & -13.27 $\pm$ 0.12 & -22.24 $\pm$ 0.07 & 6.44 $\pm$ 0.05 & 11.28 & 10.38 $\pm$ 0.03 &  8.99 $\pm$ 0.03 &  8.31 $\pm$ 0.07 &  7.92 $\pm$ 0.02 & \nodata            &   \nodata          & 0.7040 &      \\ 
2 & 15:39:28.29 & -34:46:18.3 &  \nodata            &   \nodata            & -13.60 $\pm$ 0.20 & -21.57 $\pm$ 0.12 & 6.36 $\pm$ 0.08 & 12.39 & 11.21 $\pm$ 0.02 & 10.22 $\pm$ 0.03 &  9.54 $\pm$ 0.03 &  9.06 $\pm$ 0.02 &  9.01 $\pm$ 0.13 &  8.94 $\pm$ 0.13 & 0.6347 &      \\ 
3 & 15:39:46.37 & -34:51:03.0 & \nodata             &    \nodata           & -15.25 $\pm$ 0.09 & -22.33 $\pm$ 0.05 & 6.40 $\pm$ 0.04 & 11.86 & 10.74 $\pm$ 0.03 &  9.65 $\pm$ 0.02 &  8.99 $\pm$ 0.03 &  8.84 $\pm$ 0.02 &  \nodata           &  8.64 $\pm$ 0.13 & 0.5313 &      \\ 
4 & 15:40:38.25 & -34:21:36.8 & -21.09 $\pm$ 2.53 & -30.45 $\pm$  1.74 & -30.47 $\pm$ 0.24 & -35.55 $\pm$ 0.16 & 9.09 $\pm$ 0.13 & 12.92 &            \nodata & 10.00 $\pm$ 0.02 &  9.36 $\pm$ 0.02 &  9.12 $\pm$ 0.02 &  \nodata           &  8.83 $\pm$ 0.13 & 0.4387 & con. \\ 
5 & 15:41:21.88 & -34:40:15.2 & -12.69 $\pm$ 3.07 & -23.08 $\pm$  2.98 & -19.24 $\pm$ 0.09 & -26.19 $\pm$ 0.06 & 7.28 $\pm$ 0.05 & 13.78 & 12.31 $\pm$ 0.02 & 11.09 $\pm$ 0.02 & 10.41 $\pm$ 0.02 & 10.20 $\pm$ 0.02 &  9.99 $\pm$ 0.13 &  9.95 $\pm$ 0.13 & 0.4302 &      \\ 
\enddata
\tablecomments{Sources not included in either of the member training sets are indicated in the last column by the label ``rm.".  Sources not included in the constrained member training set are indicated in the last column by the label ``con."}
\tablecomments{Table \ref{tab:members_info} is published in its entirety in the machine-readable format. A portion is shown here for guidance regarding its form and content.}
\end{splitdeluxetable*}

The properties of the (previously identified) members of the Lupus star-forming region are used by the RF to identify candidate members from the sources by searching for sources with similar properties.  
Hence, the RF is affected by the choice of labeled members (and therefore labeled desirable properties) for the training set.  
Initially, six previously identified members are removed from the training set for having measured proper motion consistent with that expected from a background source population.  
These sources are indicated in Table \ref{tab:members_info} with ``rm." in the final column.  
The remaining list of 122 members of the Lupus star-forming region covering the clouds is labeled the ``full member list".  

The ``constrained member list" accounts for possible outliers in the measured properties of the Lupus members.  
Members in the constrained member list are selected from their astrometric and photometric properties.  
Constrained members have astrometric measurements that are less than one standard deviation from any of the full membership list gaussian cluster projections (examined using the CRAN package ``mclust" \citealt{fraley18, r13}).  
Members in the constrained member list also fall to the left of the approximate lower boundary of members in the color-magnitude diagrams (see Figure \ref{fig:rf_cmds}).  
Members not included in the constrained member list only are indicated in Table \ref{tab:members_info} with ``con." in the final column.  
In total, the constrained member list consists of 106 members of the Lupus star-forming region covering the five cloud regions.  

 A literature search for stars that are spectroscopically confirmed members of Lupus in the extended catalog reveals 45 members not previously included in the on-cloud catalog.  
Tables \ref{tab:members} and \ref{tab:members_info} list these sources after the previously identified members in the on-cloud catalog.   Two of the members in the extended catalog are not included in the training set for the extended RF because their overall proper motion is consistent with a background population (see Section \ref{sec:mem}).  
The extended RF training set is drawn from the on-cloud catalog as described in Section \ref{sec:percent} so that the combined member list of 165 members (122 from the original catalog, 43 from the extended catalog) comprises of 2.5\% of the overall training set.  

\subsubsection{Non-Member List\label{sec:non_mem}}

In order to properly train the classifier, the RF training set must include both members and non-members of the Lupus star-forming region.  
There are a number of ways to identify non-members of Lupus: spectroscopically, proper motion indicative of background stars, inconsistent positions on a color-magnitude diagram, etc..  
Since the members are required to be spectroscopically confirmed, a search is conducted for sources with the same requirements for non-membership.  
A SIMBAD search for sources with spectroscopically measured spectral types that are not part of the Lupus star-forming region results in 115 sources.  
In order to better cover the wide range of property possibilities expected to be present in a diverse population of sources that are not members of the Lupus star-forming region, additional sources are labeled as non-members of Lupus from the unlabeled catalog covering the five cloud regions based on their properties.  

Sources with measured \emph{Gaia} proper motions consistent with those expected of background stars are identified as probable non-members of Lupus. 
Expected motion of a background star is \emph{Gaia} proper motion within 0.1 mas yr$^{-1}$ of a proper motion of (0,0) mas yr$^{-1}$.  
These sources are placed in the  ``background sources" list.  
Complementing this method of non-member identification, sources with proper motion inconsistent with those of the members of Lupus are also picked out.  
Sources with \emph{Gaia} proper motion greater than three standard deviations from the multivariate gaussian clusters of member proper motion (found with the CRAN package ``mclust" \citealt{fraley18, r13}) are labeled as ``inconsistent motion sources".  

Finally, sources whose position on at least one color-magnitude diagram is to the left and below the approximate lower boundary for the main trend in the color-magnitude diagrams of the members with a distance greater than 0.1 magnitudes are labeled as ``inconsistent color sources".  
In each of the possible non-member lists, ``background sources", ``inconsistent motion sources" and ``inconsistent color sources",  only  sources that have measurements for at least half of the on-cloud catalog features are retained.

\subsubsection{Within Training Set\label{sec:percent}}

Seven non-member training source lists are made to be paired with member lists to make a complete training set for the RF.   
These seven training sets can be broken up into two main groups; homogeneously selected non-members such that the two classes (members and labeled non-members) are approximately equal and mixed selections of non-members so that the member class is only a small percentage of the training set.  
The first type of training set is advantageous because it allows the RF to distinguish between different non-member lists directly and does not require any compensation for class imbalance.  
The second type of training set more accurately reflects the underlying distribution of classes in the unlabeled on-cloud catalog and covers a wider range of non-member feature properties in the training set.  

The training sets with no need to consider class imbalance (those of the first type) are called ``Spectra only", ``Background only" and ``Inconsistent proper motion".  
They are made up of the spectroscopically identified non-members, a random subsample of ``background sources" (Section \ref{sec:non_mem}) with a length equal to the full member list and a random subsample of ``inconsistent motion sources" with a length equal to twice that of the full member list respectively.   
Our smallest mixed training set, ``Mix small",  is simply the combination of the three aforementioned training sets, making the member class approximately 20\% of the full training set. 
Our final three training sets are called ``10\% mix", ``5\% mix" and ``2.5\% mix" because they are constructed from the combination of the spectroscopically identified non-members and a random sample split equally between the ``inconsistent motion sources" and the ``inconsistent color sources" such that the full member list furnishes 10\%, 5\% and 2.5\% of the total training set respectively.  

The training set data is randomly shuffled before being passed to the RF.  

\subsubsection{Feature Options\label{sec:features}}

The RF looks for the split on a node that best separates the classes given the random sample of features available.  
Different splits will be made depending on the feature chosen by the node, therefore the form of the feature (i.e. how the classifier ``sees" a feature) affects the splitting characteristic and the overall performance of the RF.   
The forms of the proper motion measurements and the photometry are tested here for optimization. 
Proper motion measurements are reported with respect to right ascension (times the cosine of the declination angle) and declination.  However, the overall shape of the distribution of members is more easily described in polar coordinates rather than Cartesian coordinates (a cluster rather than a box, see Figure \ref{fig:rf_astr}). 
Therefore, RFs are also trained on proper motion data that has been transformed to polar coordinate proper motion in magnitude and position angle.  
Proper motions from IRAC and \emph{Gaia} are used in the same form for each classifier.

Color-magnitude diagrams are used to distinguish young low-mass stars from background stars because they serve as proxies for the Hertzsprung-Russel (HR) diagram.  
The catalogs contain the photometric magnitudes for the sources, but the colors used in the color-magnitude diagrams (see Figure \ref{fig:rf_cmds}) are very specific linear combinations of those photometric bands.  
Data sets that contain only the color indicies and photometric magnitudes used in the color-magnitude diagrams are created to train the RF as an alternative to the photometric catalog data.  

\subsection{Missing Data Imputation Methods\label{sec:imputate}}

The RF cannot handle missing data points as a general rule. 
However, imputation methods can fill in the missing data holes with appropriate estimations of the values.  
The ``randomForest SRC" package offers two imputation methods: on-the-fly-imputation and missForest.

On-the-fly-imputation uses only non-missing data to make the split rule for a node when growing a decision tree in the RF.  
Once the rule has been made, the missing data holes are filled by drawing a random value from in-bag data for the feature so that the source can be sorted by the node.  
Then the source data are set back to their original data.  
This process continues until all trees are fully grown, then the missing data are imputed from the out-of-bag non-missing data from all of the trees \citep{tang17}.   
For greater accuracy, this method may be run for multiple iterations using the imputed data to grow a new RF and using the non-missing in-bag cases from that RF to re-impute the missing data \citep{ishwaran08}.  
Five different iteration values for the on-the-fly-imputation method are tested: 1, 2, 3, 5 and 10 iterations.

The missForest imputation method uses some fraction of the features (set by the user) as a multivariate response to a RF problem.  
A RF is grown on the other data (using simply the median of non-missing values to fill missing data holes if no prior imputations exist) and then used to impute the missing data within the features set aside in the beginning.  
This process is repeated until all missing data have been imputed.  
This entire process is repeated until the accuracy of the imputed data converges to a tolerance of 0.01.  
The accuracy metric for imputed data may be found in \citet{tang17}.  
feature fractions of 0.1, 0.25, 0.5, 0.75 and 1 are compared for the missForest imputation method.

\subsection{Imbalance Compensation Options\label{sec:imbalance}}

Imbalance between classes needs to be compensated for, so that the machine learning classifier optimizes class distinction rather than accuracy (which can still be high if you naively assign the majority class to all cases with a very small minority class).  
\citet{chen04} demonstrates two methods for handling class imbalance with the RF which both perform well, even for a data set with a minority class percentage of 2.3\%, class weights and balanced RF. 

Class weights are implemented such that a sample of twice the length of the member list passed to the RF is drawn from the overall training set with a probability weight of the percentage of member class in the training set for non-members and a probability weight of one minus the percentage of member class in the training set for members to grow each decision tree.  
Balanced RFs are set so that each decision tree is grown using a sample of sources drawn with replacement equal to 0.9 times the length of the number of members in the training set from both the labeled training members and the labeled training non-members.  

Of the seven training sets described in Section \ref{sec:percent} four are tested without any compensation for class imbalance; ``Spectra only", ``Background only" (no class imbalance present), ``Inconsistent proper motion" and ``Mix small".  
The use of class weights in the RF is compared for the training sets with ``Mix small" non-member lists.  
Both class weights and balanced RF methods are evaluated on ``10\% mix", ``5\% mix" and ``2.5\% mix".  
Some RFs are also trained without using a compensation for class imbalance for comparison.

\section{Comparing Random Forests\label{sec:rand_forest_comp}}

Table \ref{tab:rf} lists all of the input parameters whose combinations go into 2,240 different RFs for comparison.   

All but 42 of the member sources have at least one feature that must be imputed.  
To ensure that the imputed values for the members best reflect the properties of the member population, the imputation method utilized by the classifier is run first on just the member data, then on the non-member training set and the imputed member data set together.  The classifiers using the ``2.5\% mix" training set are trained on a random 80\% subset of the data.

\begin{deluxetable*}{lcl}
\tabletypesize{\scriptsize}
\tablecaption{Parameters for training RFs.  \label{tab:rf}}
\tablehead{
\colhead{} & \colhead{Number} & \colhead{Names} 
}
\startdata
Member List & 2 & Full member list, Constrained member list \\
Non-Member Lists & 7 & Spectra only, Background only, Inconsistent proper motion, Mix small, 10\% mix, 5\% mix, 2.5\% mix \\
feature Combinations & 4 & Measured $\mu$, Photometry, Combined $\mu$, color indices\\
Imputation Methods & 10 & On-the-fly-imputation (five iteration choices), missForest (five feature fraction choices)\\
Imbalance Compensation\tablenotemark{a} & 3 & None, Class weights, Balanced\\\hline
Total & 2,240 & RFs trained\\
\enddata
\tablenotetext{a}{Imbalance compensation is applied only to some of the RF combinations (see Section \ref{sec:imbalance} for details)}
\end{deluxetable*}

In order to compare the RFs, the classifiers are each used to predict the class labels for the full ``2.5\% mix" training set sources. This is the most fair comparison as this data set best reflects the qualities of the catalog.  Due to the bootstrapping process inherent in the RF implementation each tree within the forest is only grown on about two-thirds of the data so there is no danger of any tree over-fitting the data and the ensemble nature of the classifier protects each forest from over-fitting the training set.
Rather than return simple class labels for each source, the RF returns a ``probability" that the source is a member of the Lupus star-forming region.  
The number returned by the RF is not an actual probability of membership, but rather a probability analog.   

Four different metrics of performance are used to rank the classifiers: the F1 score, Matthews correlation coefficient (MCC), the area under the receiver operator characteristic (ROC) curve and the area under the precision recall curve.  
These metrics are calculated with the aid of the \emph{confusionMatrix} function from the CRAN package ``caret" \citep{kuhn18} in the R statistical software environment \citep{r13}. 
The F1 score and MCC change for a RF depending on the probability cutoff defined before the calculation of the metric.  
Therefore, the maximum F1 score and MCC from all possible probability cutoff values are used to rank the RFs.  
The probability cutoff values are calculated to the fourth decimal place so that the effect of changing the class label for one source in the comparison set can be evaluated.  

Table \ref{tab:rf_rank} shows the characteristics of the top six ranked RFs based on the metrics.  The ROC and the precision recall curves are built by plotting the appropriate features (true positive rate of members as a function of the false positive rate of members or the precision as a function of the true positive rate of members) for every probability cutoff value. 
Greatest weight, when ranking the classifiers, is given to the area under the curves (AUC) for the ROC curve and then the precision recall curve.  
 We use the highest-ranked RF to classify all the sources in the catalog covering the five cloud regions and find candidates.  
The ROC curve for the highest-ranked RF is 1.0000 indicating that the classifier is able to perfectly separate the members from the non-members.  
All of the top-ranked RFs are trained using the polar coordinate form for the proper motion measurements, color indices for the photometry and the 2.5\% mix non-member list. 
 The highest-ranked RF uses missForest with a feature fraction of 0.5 as its imputation method and the balanced forest method for dealing with class imbalance.

\begin{splitdeluxetable*}{cccccBccccccc}
\tabletypesize{\scriptsize}
\tablecaption{Characteristic metrics for the top six RF classifiers. \label{tab:rf_rank}}
\tablehead{
\colhead{} & \colhead{Proper Motion\tablenotemark{a}} & \colhead{Photometry\tablenotemark{b}} & \colhead{Imputation Method} & \colhead{Training Members} &  \colhead{} & \colhead{Training Non-Members} & \colhead{Class Imbalance} & \colhead{F1} & \colhead{MCC} & \colhead{ROC AUC} & \colhead{PR\tablenotemark{c} AUC}
}
\startdata
\textbf{1} & \textbf{polar} & \textbf{color indices} & \textbf{missForest, \textbf{feature} fraction 0.5} & \textbf{full member list} & \textbf{1} & \textbf{2.5\% mix} & \textbf{balanced} & \textbf{0.9959} & \textbf{0.9958} & \textbf{1.0000} & \textbf{0.9919} \\
2 & polar & color indices & missForest, feature fraction 0.25 & full member list & 2 & 2.5\% mix & balanced & 0.9919 & 0.9917 & 0.9999 & 0.9919 \\
3 & polar & color indices & on-the-fly-imputation, 2 iterations & full member list & 3 & 2.5\% mix & balanced & 0.9918 & 0.9916 & 0.9999 & 0.9917 \\
4 & polar & color indices & on-the-fly-imputation, 2 iterations & constrained member list & 4 & 2.5\% mix & balanced & 0.9837 & 0.9834 & 0.9999 & 0.9906 \\
5 & polar & color indices & on-the-fly-imputation, 3 iterations & full member list & 5 & 2.5\% mix & class weights & 0.9918 & 0.9916 & 0.9999 & 0.8195 \\
6 & polar & color indices & on-the-fly-imputation, 2 iterations & full member list & 6 & 2.5\% mix & class weights & 0.9959 & 0.9958 & 0.9999 & 0.8070\\
\enddata
\tablenotetext{a}{The form of the proper motion data used to train the RF classifier.  Form options are ``measured" for using proper motion expressed in terms of right ascension and declination and ``polar" for using proper motion in terms of magnitude and position angle.}
\tablenotetext{b}{The form of the photometry data used to train the RF classifier.  Form options are ``measured" for using only the measured photometry magnitudes and ``color indices" for using the color indices and photometry magnitudes used in color-magnitude diagrams.}
\tablenotetext{c}{Precision Recall}
\end{splitdeluxetable*}

\begin{figure*}
\epsscale{1.5}
\plotone{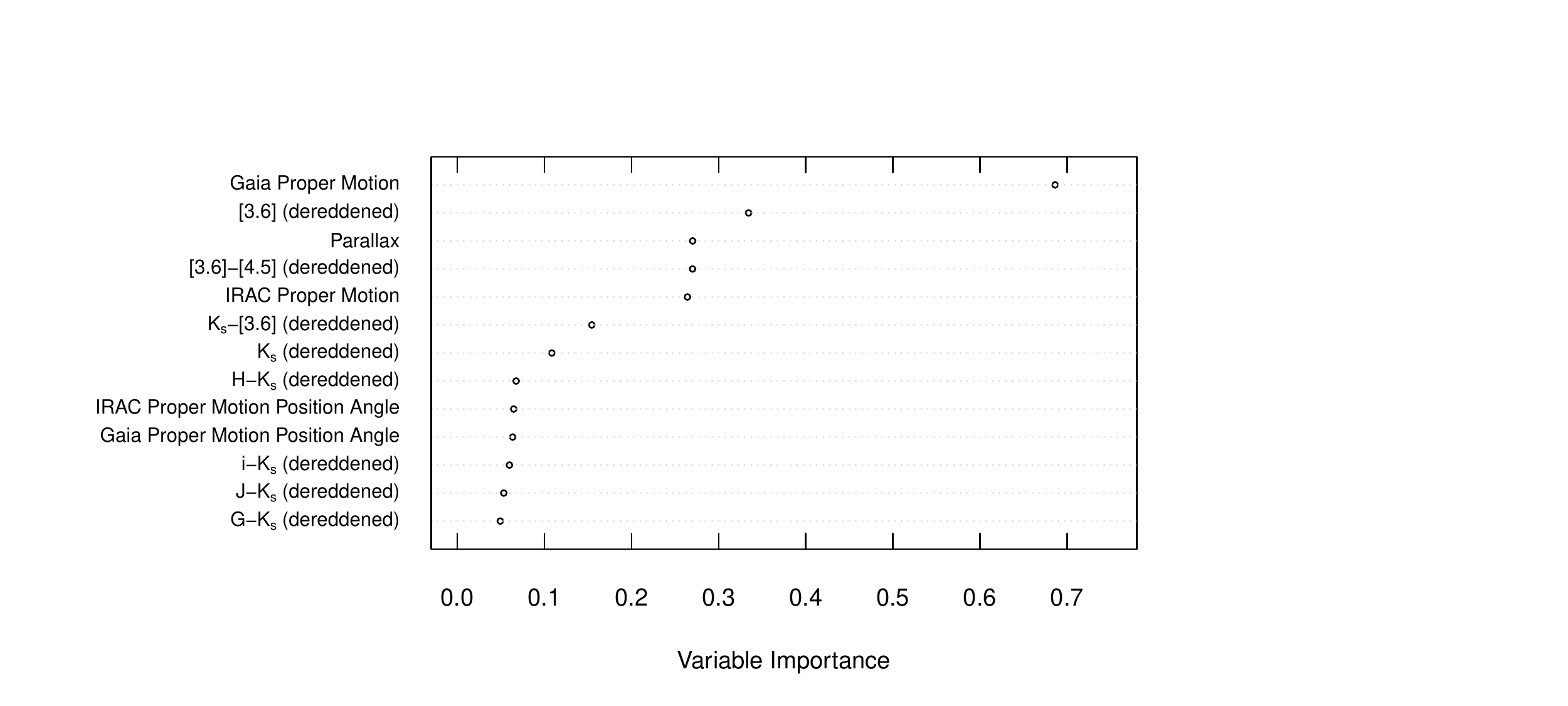}
\caption{Feature importance plot for the highest-ranked RF classifier showing the relative values of importance for each feature given to the classifier.  \label{fig:rf_var_import}}
\end{figure*}

The RF rates the importance of each feature in terms of distinguishing between classes.   
Figure \ref{fig:rf_var_import} shows that, while the magnitude of \emph{Gaia} proper motion and parallax are among the most important features, the IRAC photometry and proper motion are also very valuable in separating the classes; of the top six most important features four depend on measurements from IRAC.

\begin{figure}
\epsscale{1.2}
\plotone{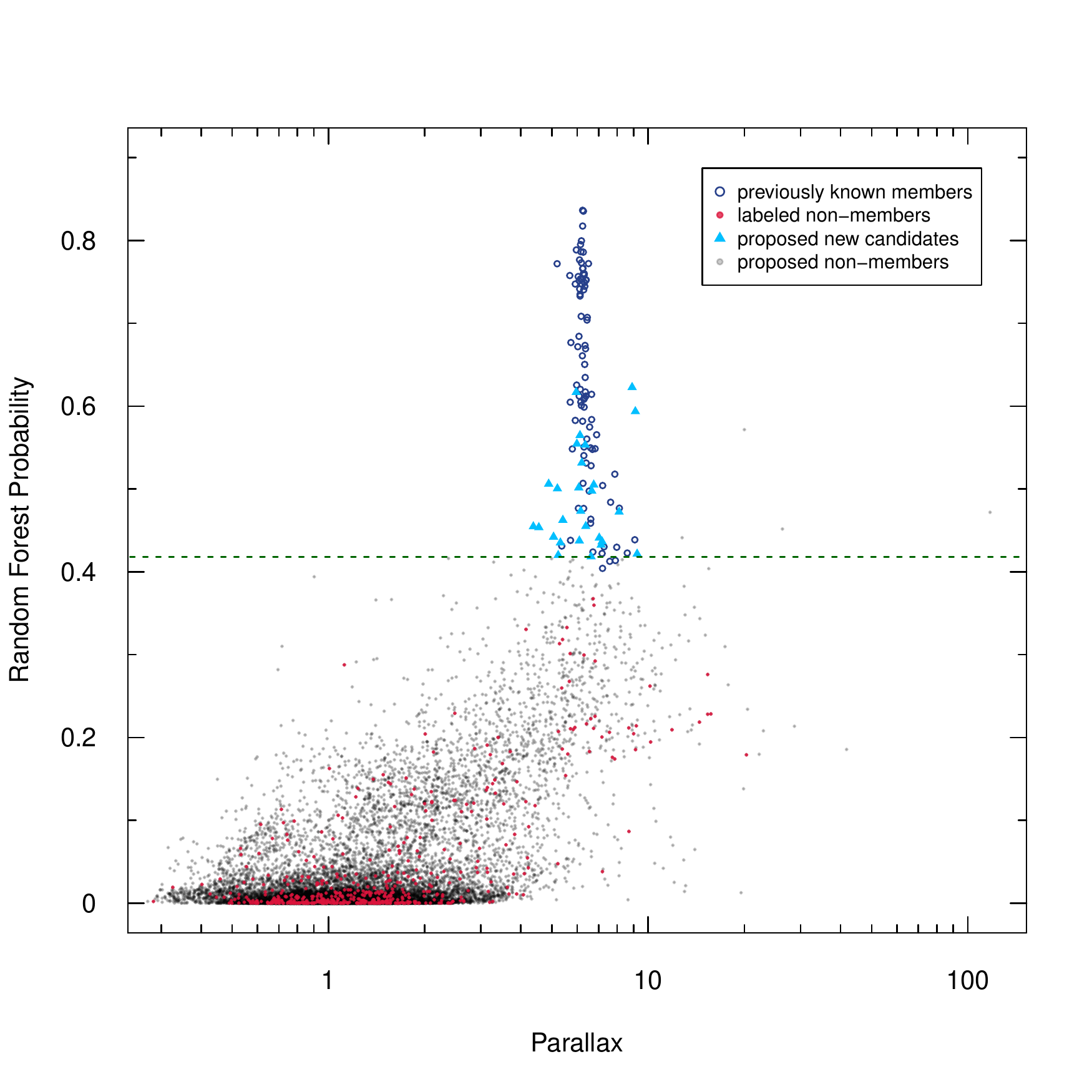}
\caption{RF probability for the on-cloud catalog of sources as a function of the \emph{Gaia} parallax (gray dots).  The previously known members (dark blue open circles) and the labeled non-members used to train the RF (red dots) are shown along with our proposed new candidate members of the Lupus star-forming region (blue triangles).  The green dashed line represents our probability threshold for determining a new candidate member.  \label{fig:rf_prob}}
\end{figure}

We choose a candidate threshold probability of 0.4180.  In the plot of RF probability as a function of parallax (Figure \ref{fig:rf_prob}), the 0.4180 threshold includes as many previously identified members as possible while excluding background sources with parallaxes significantly lower than those of previously identified members.  We remove four candidates from our list due to their large parallax values.  Table \ref{tab:conf_matrix} shows the confusion matrix for the RF using a probability threshold of 0.4180.  With a threshold of 0.4180, only three sources, all previously identified members, are mislabeled giving a true positive rate of 0.9754 and a true negative rate of 1.000.  

\begin{table}[h]
\caption{Confusion matrix for the RF classifier's training set using a probability threshold of 0.4180. \label{tab:conf_matrix}}
\noindent
\renewcommand\arraystretch{1.25}
\setlength\tabcolsep{0pt}
\begin{tabular}{l c >{\bfseries}r @{\hspace{0.5em}}c @{\hspace{0.5em}}c}
  \multirow{9}{*}{\rotatebox{90}{\parbox{0.8cm}{\bfseries\centering Predicted}}} & &
    & \multicolumn{2}{c}{\bfseries Actual Label} \\
  & \phantom{'}& & \bfseries m & \bfseries n\\
  & & m$'$ & \MyBox{119} & \MyBox{0} \\[1.5em]
  & & n$'$ & \MyBox{3} & \MyBox{4873}\\
\end{tabular}

\end{table}

Figure \ref{fig:rf_prob} clearly shows that the number of sources with parallax measurements less than that of the members of the Lupus star-forming region (open blue circles) greatly outweigh those with parallax measurements greater than the members; i.e. there are more background sources in the on-cloud catalog than sources foreground to the Lupus star-forming region.  
In the entire on-cloud catalog of 716,530 sources, 66 have a measured parallax greater than 10 mas (only five of which are labeled as non-members in the RF classifier's training set) compared to a population of 13,926 sources with parallaxes less than 2 mas. 
It is not surprising that the classifier has a more difficult time determining a more realistic probability for sources with large parallax and so we remove those sources from our candidate list.  
A parallax of 10 mas corresponds to a distance of 100 pc. 
The smallest distance to previously identified members is 110.1 pc, is almost 40 pc shorter than the mean distance to the members of the Lupus star-forming region, therefore a cut-off of 100 pc is reasonable.

\section{Candidate New Members in the Five Lupus Clouds \label{sec:cands}}

\begin{figure*}
\gridline{\fig{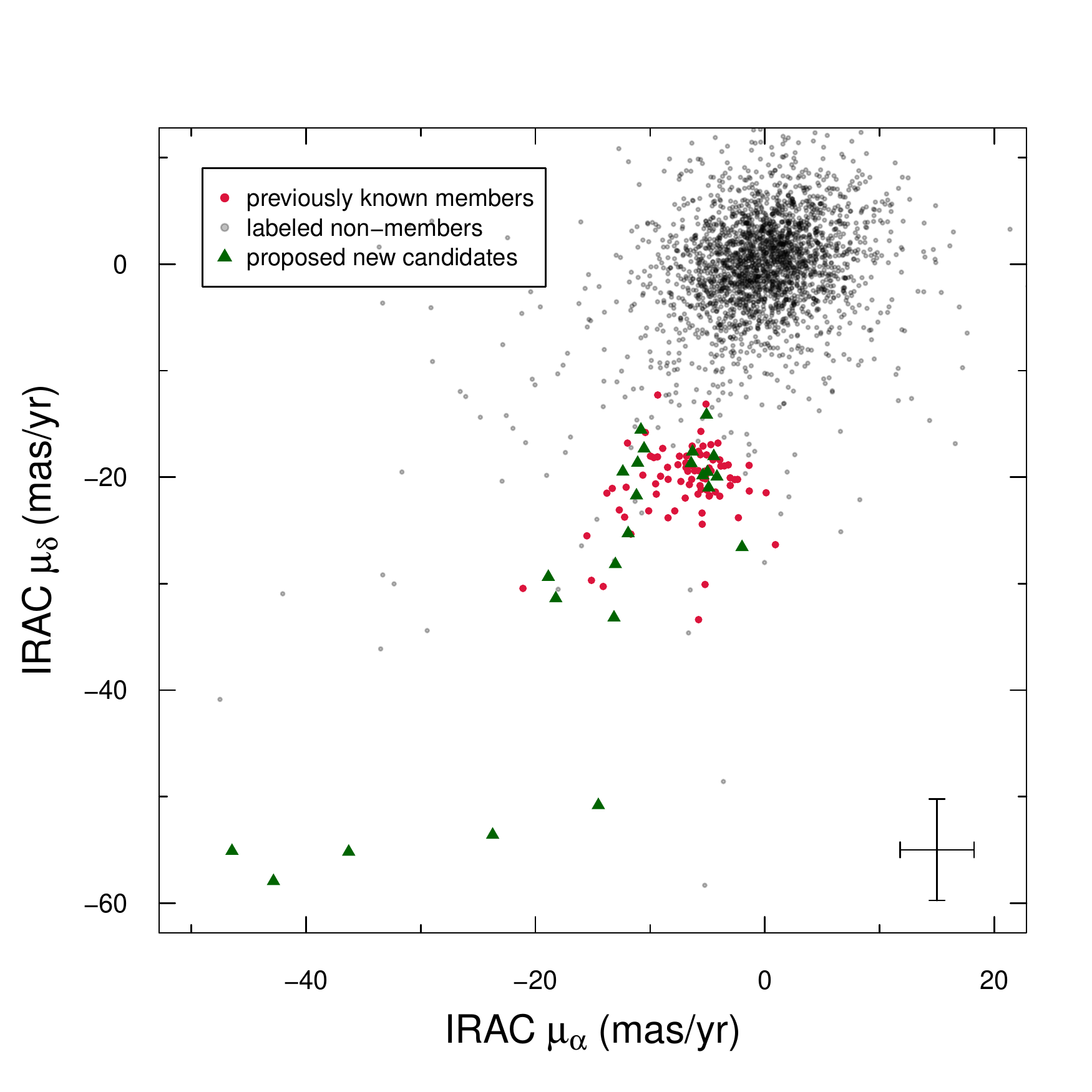}{0.5\textwidth}{(a)}
\fig{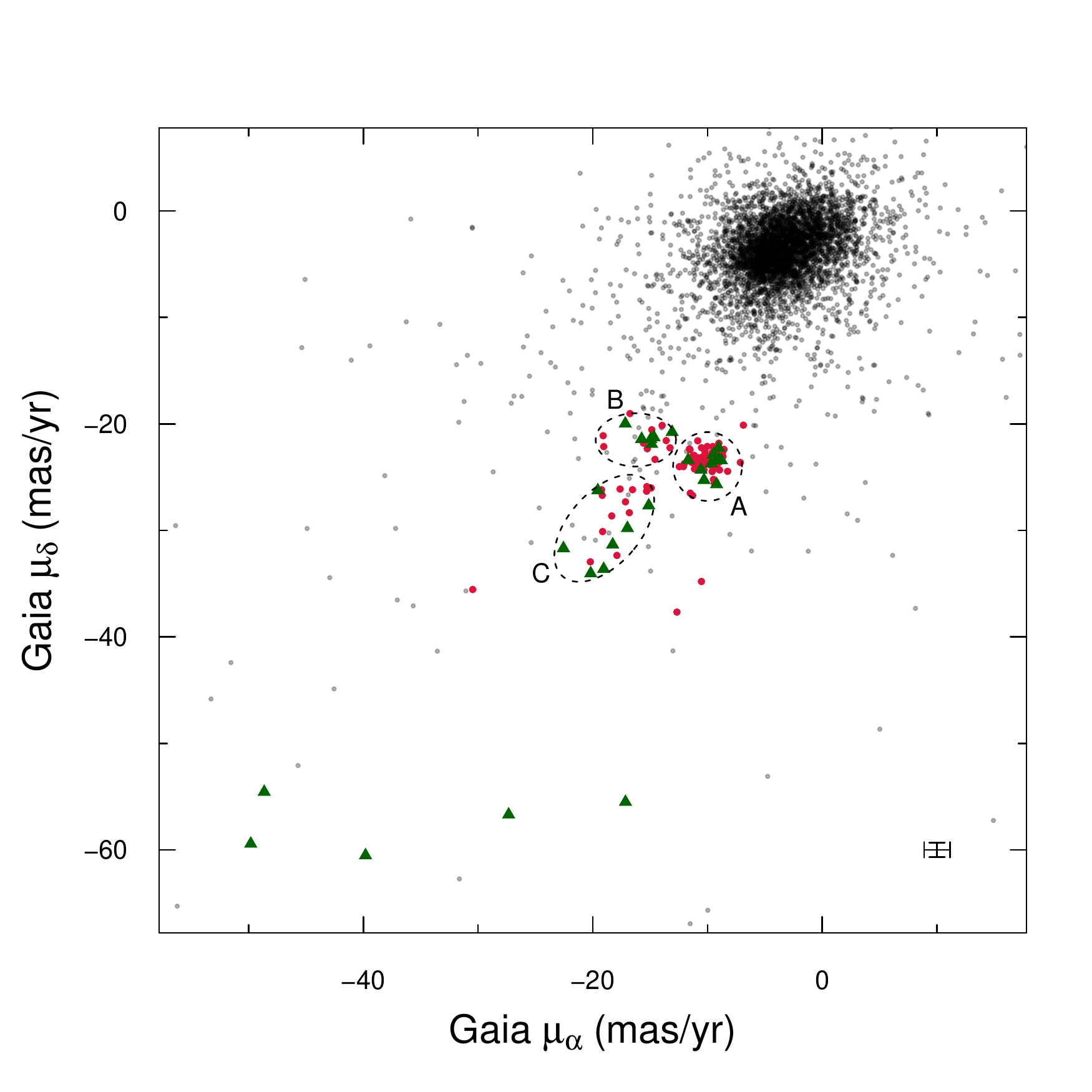}{0.5\textwidth}{(b)}
}
\gridline{\fig{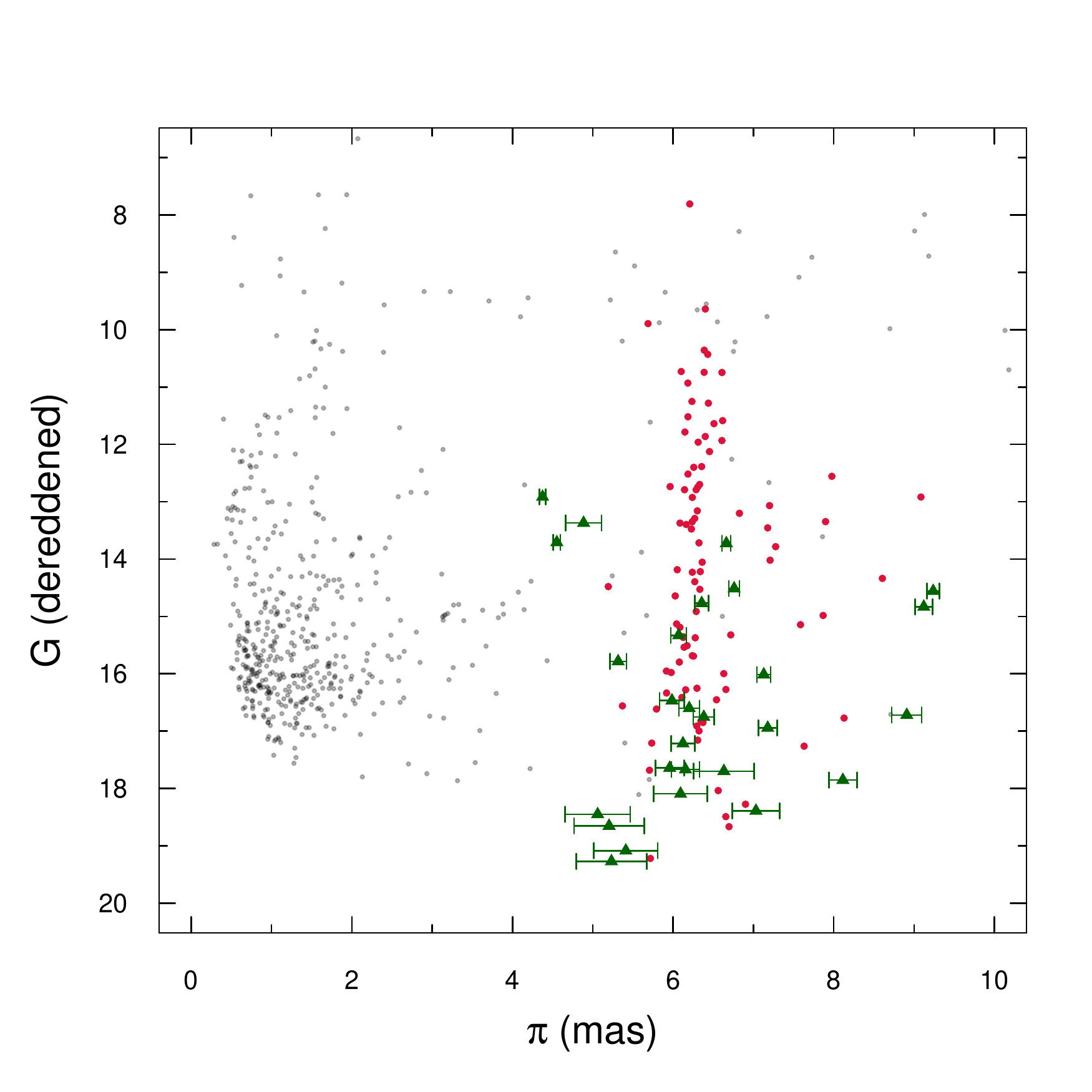}{0.5\textwidth}{(c)}
}
\caption{Proper motion and parallax plots for the 2.5\% mix training set (previously identified members are red dots, training set non-members are gray dots) and the proposed candidate members of Lupus  (green triangles) identified by our highest-ranked RF.  The mean error bars in proper motion are shown in the bottom left-hand corner of plots (a) and (b).  Plot (b) also shows three labeled proper motion groups; A, B and C are discussed in the text.   \label{fig:rf_astr}}
\end{figure*}

\begin{figure*}
\gridline{\fig{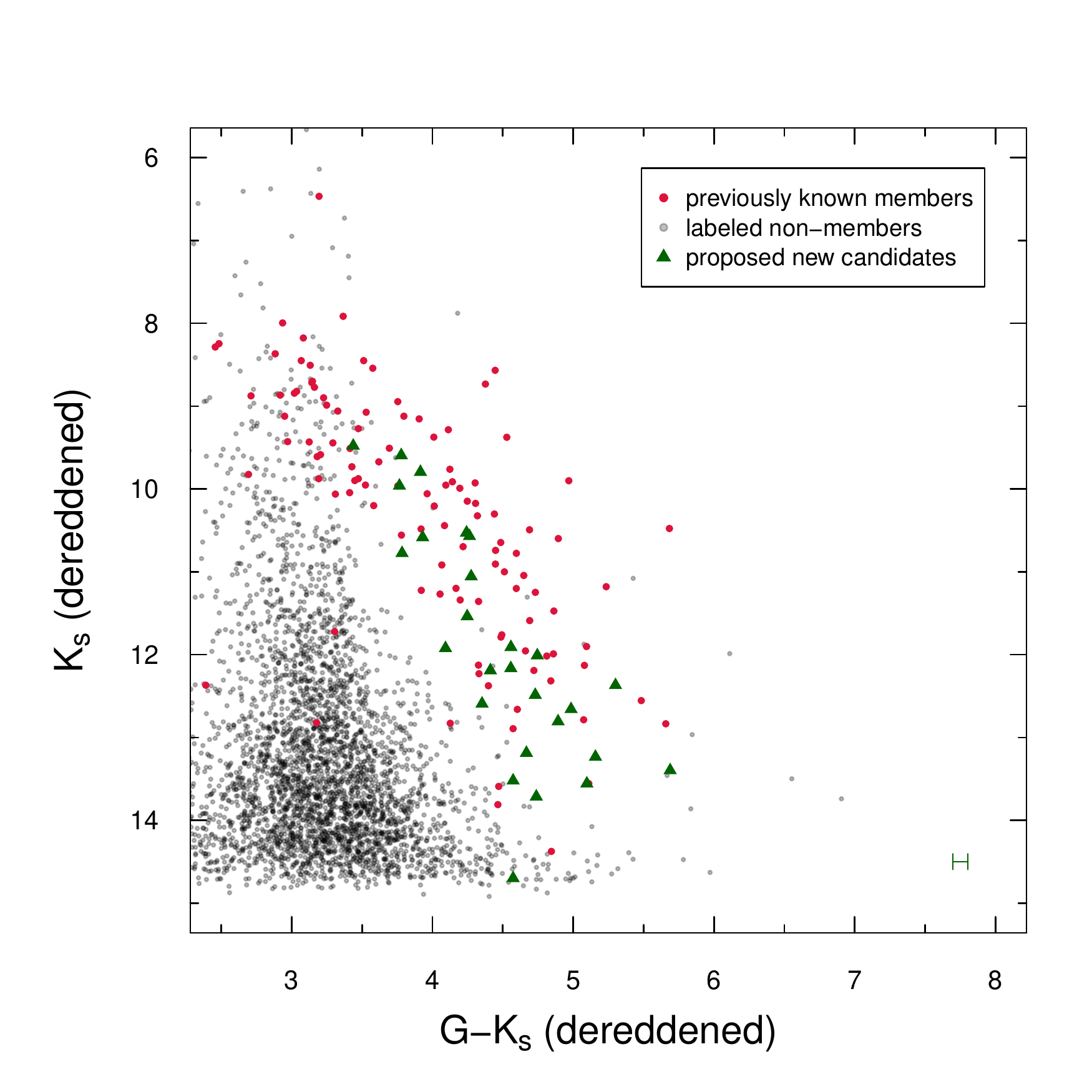}{0.5\textwidth}{(a)}
\fig{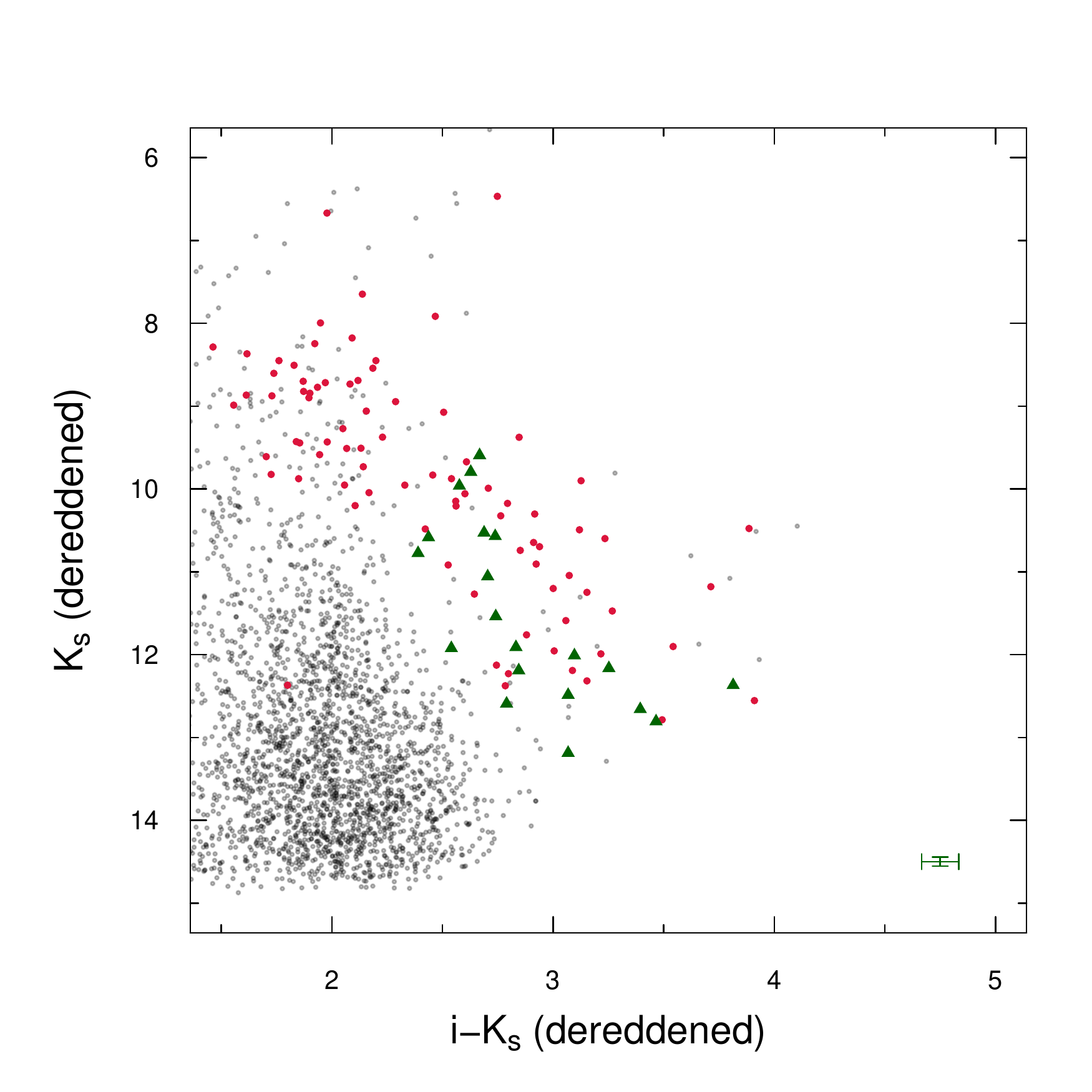}{0.5\textwidth}{(b)}
}
\gridline{\fig{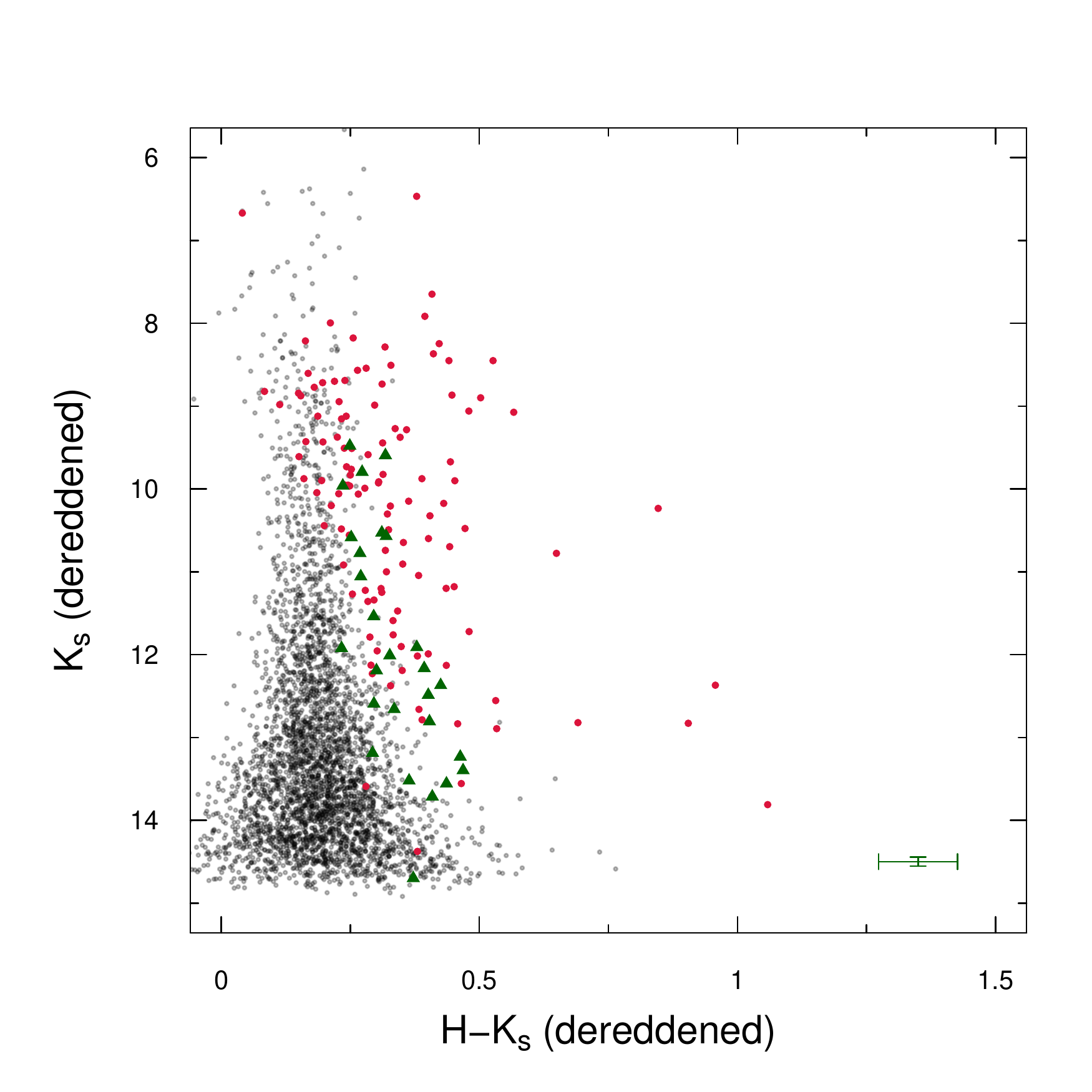}{0.5\textwidth}{(c)}
\fig{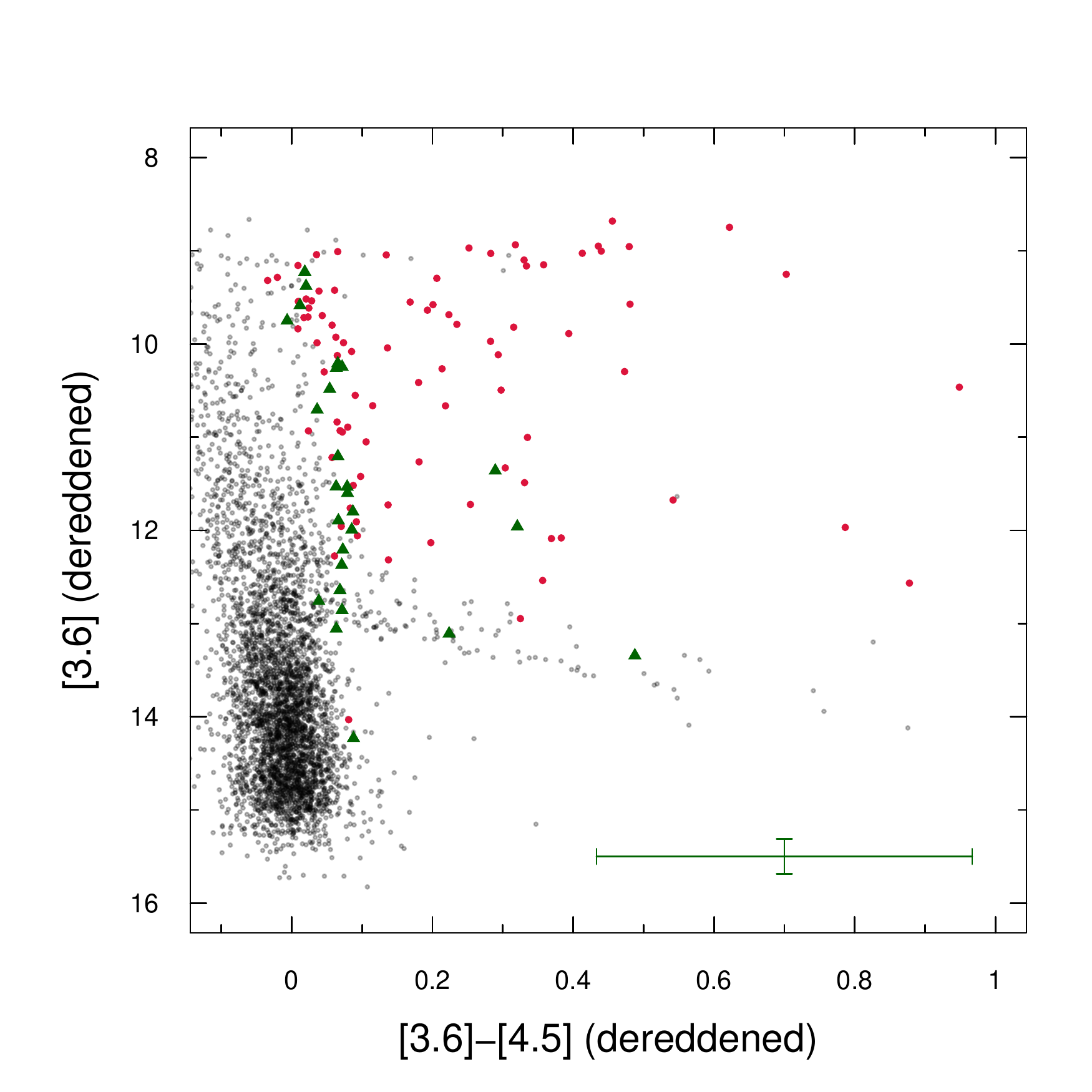}{0.5\textwidth}{(d)}
}
\caption{Extinction-corrected color-magnitude diagrams for the 2.5\% mix training set (members are red dots, training set non-members are gray dots) and the proposed candidate members of Lupus (green triangles) identified by our highest-ranked RF.  The mean error bars for the points are plotted in the bottom right-hand corner of each plot.  \label{fig:rf_cmds}}
\end{figure*}

The 27 candidate new members in the five Lupus cloud regions are superbly isolated by the classifier from field stars and have properties similar in most respects to the confirmed members.   Tables \ref{tab:cands_astrometry} and \ref{tab:cands_photometry} list the on-cloud catalog information for our candidate members in the Lupus clouds. Figures \ref{fig:rf_astr} and \ref{fig:rf_cmds} compare the astrometry of the on-cloud candidates and the previously identified members and the positions of the on-cloud candidates and the previously identified members on color-magnitude diagrams respectively.  

\begin{deluxetable*}{ccccccccl}
\tabletypesize{\scriptsize}
\tablecaption{Astrometry information for the candidates identified by the highest-ranked RF with a probability threshold of 0.4180. \label{tab:cands_astrometry}}
\tablehead{
\colhead{} & \colhead{RA} & \colhead{DEC} &  \colhead{$\mu_{\alpha}$ (IRAC)} &  \colhead{$\mu_{\delta}$ (IRAC)} & \colhead{$\mu_{\alpha}$ (\emph{Gaia})} &  \colhead{$\mu_{\delta}$ (\emph{Gaia})}  & \colhead{Parallax} & \colhead{SIMBAD}\\
\colhead{} & \colhead{hh:mm:ss.ss} & \colhead{dd:mm:ss.ss}  & \colhead{mas yr$^{-1}$} & \colhead{mas yr$^{-1}$} & \colhead{mas yr$^{-1}$} & \colhead{mas yr$^{-1}$} & \colhead{mas} & \colhead{Name}
}
\startdata
1 & 15:39:14.91 & -33:28:47.37 & -46.4 $\pm$ 3.5 & -55.1 $\pm$ 4.2 & -48.64 $\pm$ 0.23 & -54.56 $\pm$ 0.16 & 4.89 $\pm$ 0.22 &  \\
2 & 15:39:21.79 & -34:0:19.67 &    \nodata                &       \nodata         & -17.17 $\pm$ 0.49 & -19.97 $\pm$ 0.37 & 6.63 $\pm$ 0.38 & 2MASS J15392180-3400195\\
3 & 15:59:4.58 & -42:10:58.60 &  \nodata                  &    \nodata            & -10.56 $\pm$ 0.30 & -24.31 $\pm$ 0.22 & 6.15 $\pm$ 0.17 & 2MASS J15590456-4210583\\
4 & 15:59:44.11 & -42:11:55.97 & -6.4 $\pm$ 2.6 & -18.7 $\pm$ 4.5 & -9.02 $\pm$ 0.17 & -22.25 $\pm$ 0.13 & 6.07 $\pm$ 0.10 &  \\
5 & 16:0:41.04 & -41:24:46.15 & -11.2 $\pm$ 1.3 & -21.7 $\pm$ 2.1 & -15.12 $\pm$ 0.18 & -27.66 $\pm$ 0.12 & 7.13 $\pm$ 0.09 &  \\
6 & 16:1:42.84 & -41:27:28.51 & -23.7 $\pm$ 4.3 & -53.6 $\pm$ 4.7 & -27.33 $\pm$ 1.08 & -56.69 $\pm$ 0.63 & 5.23 $\pm$ 0.44 &  \\
7 & 16:6:58.71 & -39:4:5.26 &    \nodata          &    \nodata            & -11.69 $\pm$ 0.26 & -23.39 $\pm$ 0.16 & 6.38 $\pm$ 0.13 & 2MASS J16065870-3904051\\
8 & 16:8:17.40 & -39:1:6.29 & -5.0 $\pm$ 0.8 & -19.5 $\pm$ 1.0 & -9.50 $\pm$ 0.15 & -22.90 $\pm$ 0.10 & 6.36 $\pm$ 0.08 & 2MASS J16081739-3901062\\
9 & 16:9:8.47 & -39:2:13.84 & -36.3 $\pm$ 1.5 & -55.2 $\pm$ 1.3 & -39.81 $\pm$ 0.08 & -60.52 $\pm$ 0.05 & 4.37 $\pm$ 0.04 & 2MASS J16090849-3902133\\
10 & 16:10:10.19 & -38:52:23.23 & -4.2 $\pm$ 0.4 & -20.0 $\pm$ 0.2 & -8.79 $\pm$ 0.87 & -23.42 $\pm$ 0.50 & 5.41 $\pm$ 0.40 &  \\
11 & 16:17:18.10 & -36:46:30.84 & -6.3 $\pm$ 4.9 & -17.6 $\pm$ 2.7 & -9.03 $\pm$ 0.36 & -23.32 $\pm$ 0.20 & 5.96 $\pm$ 0.18 & 2MASS J16171811-3646306\\
12 & 16:17:24.86 & -36:57:40.78 & -2.0 $\pm$ 4.6 & -26.6 $\pm$ 2.8 & -9.21 $\pm$ 0.14 & -25.68 $\pm$ 0.08 & 6.76 $\pm$ 0.07 & 2MASS J16172485-3657405\\
13 & 16:17:26.59 & -37:15:58.93 & -13.0 $\pm$ 0.8 & -28.2 $\pm$ 3.1 & -18.27 $\pm$ 0.38 & -31.33 $\pm$ 0.23 & 8.11 $\pm$ 0.17 &  \\
14 & 16:17:55.59 & -37:38:21.82 & -11.9 $\pm$ 3.6 & -25.3 $\pm$ 4.8 & -16.98 $\pm$ 0.25 & -29.79 $\pm$ 0.14 & 7.18 $\pm$ 0.12 &  \\
15 & 16:18:23.22 & -37:30:38.72 & -5.4 $\pm$ 5.4 & -19.8 $\pm$ 5.4 & -10.30 $\pm$ 0.31 & -25.27 $\pm$ 0.17 & 6.12 $\pm$ 0.15 &  \\
16 & 16:19:26.82 & -36:51:23.97 & -13.2 $\pm$ 6.9 & -33.2 $\pm$ 5.0 & -20.18 $\pm$ 0.33 & -34.04 $\pm$ 0.24 & 8.91 $\pm$ 0.19 & 2MASS J16192684-3651235\\
17 & 16:19:54.81 & -37:40:24.70 & -18.9 $\pm$ 6.7 & -29.4 $\pm$ 5.1 & -22.56 $\pm$ 0.09 & -31.68 $\pm$ 0.07 & 4.55 $\pm$ 0.05 &  \\
18 & 16:20:34.28 & -37:40:24.98 & -11.1 $\pm$ 5.0 & -18.7 $\pm$ 6.1 & -19.56 $\pm$ 0.60 & -26.24 $\pm$ 0.50 & 7.03 $\pm$ 0.29 &  \\
19 & 16:21:33.80 & -41:8:37.31 & -10.8 $\pm$ 5.6 & -15.5 $\pm$ 4.7 & -14.95 $\pm$ 0.90 & -21.42 $\pm$ 0.50 & 5.20 $\pm$ 0.44 &  \\
20 & 16:21:35.75 & -41:2:45.29 & -5.1 $\pm$ 4.0 & -14.1 $\pm$ 8.7 & -15.74 $\pm$ 0.68 & -21.42 $\pm$ 0.37 & 6.09 $\pm$ 0.33 &  \\
21 & 16:21:39.85 & -37:0:58.95 & -18.2 $\pm$ 5.6 & -31.4 $\pm$ 2.7 & -19.05 $\pm$ 0.20 & -33.63 $\pm$ 0.15 & 9.12 $\pm$ 0.11 &  \\
22 & 16:22:31.09 & -41:5:27.43 & -12.4 $\pm$ 6.0 & -19.5 $\pm$ 8.6 & -14.66 $\pm$ 0.26 & -21.27 $\pm$ 0.16 & 5.99 $\pm$ 0.15 &  \\
23 & 16:23:51.26 & -39:59:21.23 & -4.9 $\pm$ 3.5 & -21.0 $\pm$ 4.5 & -9.67 $\pm$ 0.13 & -23.72 $\pm$ 0.09 & 6.66 $\pm$ 0.05 &  \\
24 & 16:24:24.86 & -39:48:18.77 & -4.5 $\pm$ 2.9 & -18.0 $\pm$ 5.8 & -13.09 $\pm$ 0.22 & -20.77 $\pm$ 0.16 & 5.32 $\pm$ 0.10 &  \\
25 & 16:25:58.03 & -40:2:23.45 & -14.5 $\pm$ 4.8 & -50.8 $\pm$ 5.9 & -17.15 $\pm$ 0.14 & -55.51 $\pm$ 0.11 & 9.24 $\pm$ 0.08 &  \\
26 & 16:26:10.83 & -40:5:18.63 & -42.8 $\pm$ 5.2 & -57.9 $\pm$ 11.7 & -49.81 $\pm$ 0.82 & -59.43 $\pm$ 0.54 & 5.06 $\pm$ 0.41 &  \\
27 & 16:26:54.38 & -40:1:52.38 & -10.5 $\pm$ 4.3 & -17.3 $\pm$ 4.1 & -14.87 $\pm$ 0.23 & -21.88 $\pm$ 0.18 & 6.20 $\pm$ 0.13 & \\
\enddata
\tablecomments{Candidates that have been examined by other studies have their name in the rightmost column. }
\tablecomments{Table \ref{tab:cands_astrometry} is also available in machine readable format.}
\end{deluxetable*}

\begin{deluxetable*}{ccccccccll}
\tabletypesize{\scriptsize}
\tablecaption{Photometry information for the candidates identified by the highest ranked RF with a probability threshold of 0.4180.  \label{tab:cands_photometry}}
\tablehead{
\colhead{} & \colhead{$G$} & \colhead{\emph{i}} & \colhead{$J$} & \colhead{$H$} & \colhead{$K_{s}$} & \colhead{[3.6]} & \colhead{[4.5]} & \colhead{P$_{RF}$\tablenotemark{a}} & \colhead{Disk?} \\
\colhead{} & \colhead{mag.}  & \colhead{mag.} & \colhead{mag.} & \colhead{mag.} & \colhead{mag.} & \colhead{mag.} & \colhead{mag.} & \colhead{} & \colhead{}
}
\startdata
1 & 13.37 & 12.26 $\pm$ 0.03 & 10.59 $\pm$ 0.02 & 9.91 $\pm$ 0.02 & 9.59 $\pm$ 0.02 & 9.38 $\pm$ 0.13 & 9.36 $\pm$ 0.13 & 0.5062 & N \\
2 & 17.70 & 16.27 $\pm$ 0.07 & 13.89 $\pm$ 0.02 & 13.21 $\pm$ 0.02 & 12.81 $\pm$ 0.03 & 12.37 $\pm$ 0.13 & 12.30 $\pm$ 0.13 & 0.4183 & N \\
3 & 17.67 & 16.18 $\pm$ 0.06 & 13.43 $\pm$ 0.02 & 12.79 $\pm$ 0.02 & 12.37 $\pm$ 0.02 & 11.89 $\pm$ 0.13 & 11.83 $\pm$ 0.13 & 0.4733 & N \\
4 & 15.33 & 13.76 $\pm$ 0.03 & 11.95 $\pm$ 0.02 & 11.33 $\pm$ 0.02 & 11.06 $\pm$ 0.02 & 10.70 $\pm$ 0.13 & 10.67 $\pm$ 0.13 & 0.5017 & N \\
5 & 16.01 & 14.46 $\pm$ 0.03 & 12.78 $\pm$ 0.02 & 12.16 $\pm$ 0.02 & 11.92 $\pm$ 0.03 & 11.53 $\pm$ 0.13 & 11.47 $\pm$ 0.13 & 0.4324 & N \\
6 & 19.27 & \nodata            & 15.57 $\pm$ 0.06 & 15.07 $\pm$ 0.09 & 14.70 $\pm$ 0.08 & 14.23 $\pm$ 0.14 & 14.14 $\pm$ 0.13 & 0.4200 & N \\
7 & 16.75 & 15.11 $\pm$ 0.04 & 12.95 $\pm$ 0.02 & 12.34 $\pm$ 0.03 & 12.01 $\pm$ 0.03 & 11.60 $\pm$ 0.13 & 11.52 $\pm$ 0.13 & 0.4550 & N \\
8 & 14.77 & 13.22 $\pm$ 0.03 & 11.46 $\pm$ 0.02 & 10.84 $\pm$ 0.02 & 10.53 $\pm$ 0.02 & 10.26 $\pm$ 0.13 & 10.19 $\pm$ 0.13 & 0.5532 & N \\
9 & 12.92 &  \nodata             & 10.33 $\pm$ 0.02 & 9.73 $\pm$ 0.02 & 9.48 $\pm$ 0.02 & 9.23 $\pm$ 0.13 & 9.21 $\pm$ 0.13 & 0.4547 & N \\
10 & 19.09 &   \nodata           & 14.55 $\pm$ 0.04 & 13.87 $\pm$ 0.04 & 13.40 $\pm$ 0.05 & 12.86 $\pm$ 0.13 & 12.78 $\pm$ 0.18 & 0.4625 & N \\
11 & 17.64 & 16.05 $\pm$ 0.06 & 13.66 $\pm$ 0.02 & 12.99 $\pm$ 0.02 & 12.66 $\pm$ 0.03 & 11.96 $\pm$ 0.13 & 11.64 $\pm$ 0.13 & 0.6168 & Y \\
12 & 14.51 & 13.02 $\pm$ 0.03 & 11.47 $\pm$ 0.02 & 10.84 $\pm$ 0.03 & 10.58 $\pm$ 0.02 & 10.24 $\pm$ 0.13 & 10.17 $\pm$ 0.13 & 0.5051 & N \\
13 & 17.85 & 16.26 $\pm$ 0.07 & 14.06 $\pm$ 0.02 & 13.48 $\pm$ 0.04 & 13.19 $\pm$ 0.03 & 12.64 $\pm$ 0.13 & 12.57 $\pm$ 0.13 & 0.4723 & N \\
14 & 16.94 & 15.38 $\pm$ 0.05 & 13.45 $\pm$ 0.03 & 12.89 $\pm$ 0.03 & 12.59 $\pm$ 0.03 & 12.21 $\pm$ 0.13 & 12.14 $\pm$ 0.13 & 0.4365 & N \\
15 & 17.22 & 15.56 $\pm$ 0.07 & 13.44 $\pm$ 0.03 & 12.89 $\pm$ 0.02 & 12.49 $\pm$ 0.02 & 11.99 $\pm$ 0.13 & 11.91 $\pm$ 0.13 & 0.5648 & N \\
16 & 16.72 & 15.42 $\pm$ 0.05 & 13.17 $\pm$ 0.03 & 12.56 $\pm$ 0.03 & 12.16 $\pm$ 0.03 & 11.36 $\pm$ 0.13 & 11.07 $\pm$ 0.13 & 0.6228 & Y \\
17 & 13.71 & 12.42 $\pm$ 0.02 & 10.71 $\pm$ 0.02 & 10.07 $\pm$ 0.02 & 9.79 $\pm$ 0.02 & 9.58 $\pm$ 0.13 & 9.57 $\pm$ 0.13 & 0.4537 & N \\
18 & 18.39 &    \nodata          & 14.25 $\pm$ 0.03 & 13.70 $\pm$ 0.04 & 13.23 $\pm$ 0.04 & 12.76 $\pm$ 0.14 & 12.72 $\pm$ 0.13 & 0.4410 & N \\
19 & 18.65 &     \nodata         & 14.67 $\pm$ 0.05 & 13.99 $\pm$ 0.05 & 13.56 $\pm$ 0.05 & 13.11 $\pm$ 0.13 & 12.88 $\pm$ 0.23 & 0.5005 & Y \\
20 & 18.09 &  \nodata            & 14.57 $\pm$ 0.05 & 13.89 $\pm$ 0.06 & 13.52 $\pm$ 0.05 & 13.05 $\pm$ 0.13 & 12.99 $\pm$ 0.13 & 0.4375 & N \\
21 & 14.83 & 13.31 $\pm$ 0.03 & 11.46 $\pm$ 0.02 & 10.89 $\pm$ 0.02 & 10.57 $\pm$ 0.02 & 10.21 $\pm$ 0.13 & 10.14 $\pm$ 0.13 & 0.5937 & N \\
22 & 16.47 & 14.74 $\pm$ 0.05 & 12.97 $\pm$ 0.02 & 12.29 $\pm$ 0.03 & 11.91 $\pm$ 0.02 & 11.53 $\pm$ 0.13 & 11.45 $\pm$ 0.14 & 0.5545 & N \\
23 & 13.73 & 12.54 $\pm$ 0.05 & 10.87 $\pm$ 0.02 & 10.20 $\pm$ 0.02 & 9.96 $\pm$ 0.02 & 9.75 $\pm$ 0.13 & 9.75 $\pm$ 0.13 & 0.4977 & N \\
24 & 15.78 & 14.28 $\pm$ 0.04 & 12.51 $\pm$ 0.03 & 11.83 $\pm$ 0.03 & 11.54 $\pm$ 0.03 & 11.20 $\pm$ 0.13 & 11.14 $\pm$ 0.13 & 0.4352 & N \\
25 & 14.56 & 13.16 $\pm$ 0.02 & 11.63 $\pm$ 0.02 & 11.04 $\pm$ 0.02 & 10.77 $\pm$ 0.02 & 10.48 $\pm$ 0.13 & 10.43 $\pm$ 0.13 & 0.4217 & N \\
26 & 18.45 &   \nodata           & 14.80 $\pm$ 0.05 & 14.12 $\pm$ 0.05 & 13.71 $\pm$ 0.06 & 13.34 $\pm$ 0.13 & 12.86 $\pm$ 0.22 & 0.4422 & Y \\
27 & 16.60 & 15.04 $\pm$ 0.05 & 13.10 $\pm$ 0.03 & 12.49 $\pm$ 0.03 & 12.19 $\pm$ 0.03 & 11.80 $\pm$ 0.13 & 11.71 $\pm$ 0.15 & 0.5315 & N \\ 
\enddata
\tablenotetext{a}{Probability assigned to source by the RF Classifier.}
\tablecomments{The likely presence of a disk as determined in Section \ref{sec:cand_disk} is indicated in the last column. }
\tablecomments{Table \ref{tab:cands_photometry} is also available in machine readable format.}
\end{deluxetable*}

The three plots in Figure \ref{fig:rf_astr} compare the astrometric properties of the members and candidates to the labeled non-members of the training set.  Most of the on-cloud candidates show the same distribution in proper motion as the previously identified members.  Figures \ref{fig:rf_astr} (a) and (b) show that of the 27 on-cloud candidates, all but five of them lie in the same kinematic space as the previously identified members.  The five discrepant candidates are candidates 1, 6, 9, 25 and 26.  Candidates 6 and 26 are very faint sources (see Table \ref{tab:cands_photometry}) with \emph{G} $> 18$ mag.  Because of their overall faint magnitudes and high proper motion as compared to the previously identified members, these stars may be contaminant field stars or older M stars dynamically ejected from their birth locations.  The on-cloud candidate members have a slightly larger spread in parallax than the previously identified members, indicating that the depth of the Lupus clouds might be larger than previously known (see Appendix \ref{sec:dist_clouds} for further discussion).

Figure \ref{fig:rf_cmds} shows the positions of the training set non-members, previously identified members and our candidates on four different color-magnitude diagrams.  The candidates' positions on those diagrams overlap with the members in all of the plots and are generally well separated from the labeled non-member sources.  
This further demonstrates the ability of our RF to distinguish sources in the Lupus star-forming region from field stars.  
Note that, while the photometry measurements have been dereddened to correct for extinction effects where possible (see Section \ref{sec:phot}), only approximately 25\% of the sources in the on-cloud catalog are able to be corrected for extinction effects; the rest of the sources are plotted as measured by the proper photometric survey.  

Eight of our candidates have been previously examined in other studies of Lupus as candidate members and are labeled in Table \ref{tab:cands_astrometry}.
Additional information on the previously examined candidates can be found in the Appendix \ref{sec:named_cands}.  

\section{Extending the Search\label{sec:ext}}

The RF chosen in Section \ref{sec:rand_forest_comp} relies heavily on IRAC data when classifying sources.  Of the five features involving IRAC measurements (out of 13 total) in Figure \ref{fig:rf_var_import}, four of them rank in the top six most important features for the RF.  If this RF  were to be applied to the extended catalog data, the values for all of the features including IRAC data would have to be imputed using only the IRAC values in the small sample of ~5,000 sources from the training set to guide the imputation.  With no IRAC feature values in the extended data set, these values would not be very useful in classifying the sources.  Therefore, we train a new RF  (the extended RF) using only the data available in the extended catalog.  The extended RF was created to have the best performing tuning parameters from Section \ref{sec:rand_forest_comp}.  
 
\begin{figure*}
\epsscale{0.5}
\plotone{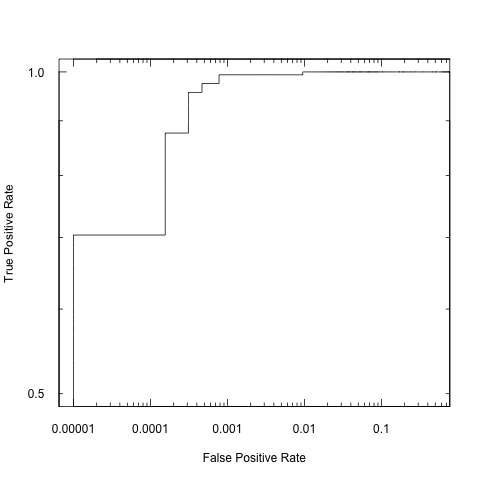}
\caption{The ROC curve (a) and the precision recall curve (b) for the RF trained for classifying the extended catalog. \label{fig:auc_plots_ext}}
\end{figure*}

\begin{figure*}
\epsscale{1.25}
\plotone{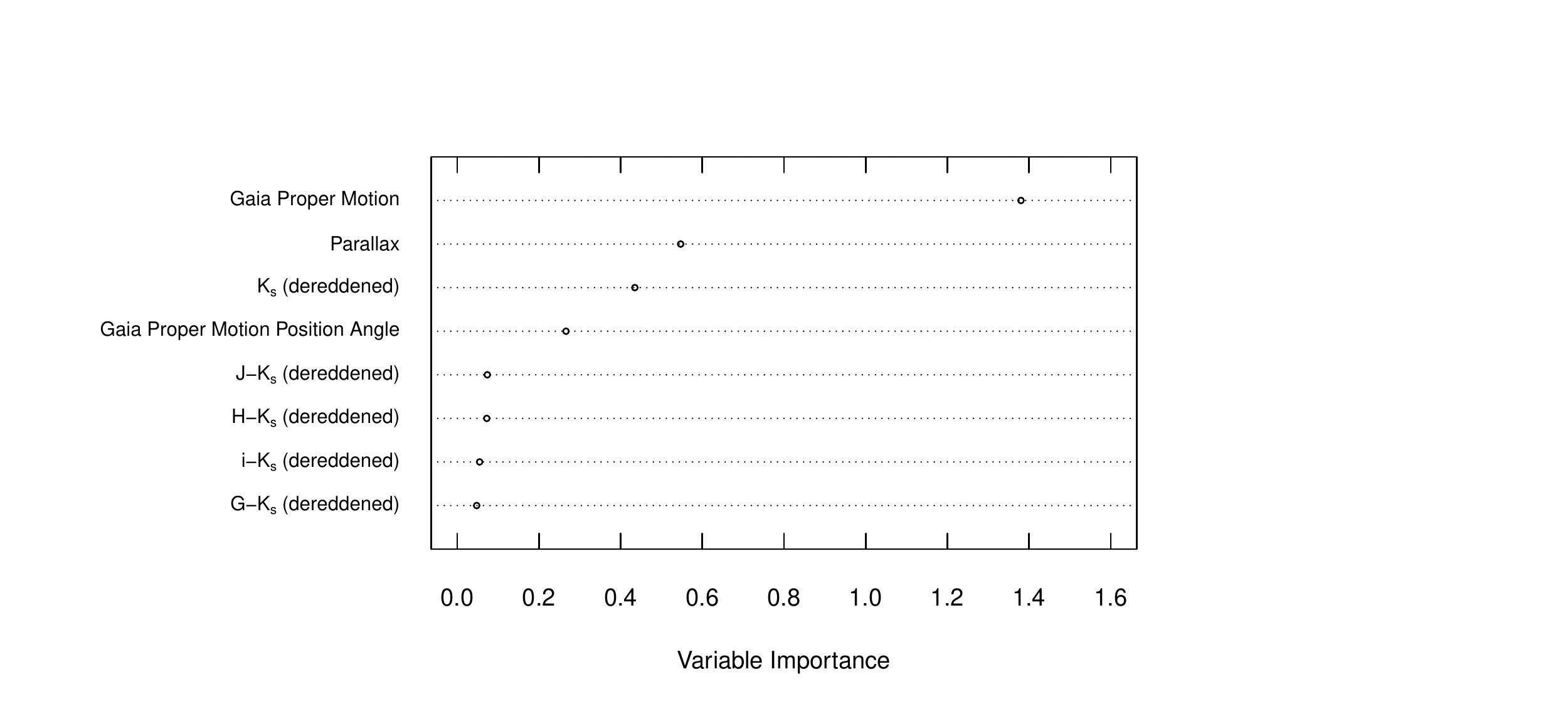}
\caption{Feature importance plot for the RF trained for classifying the extended catalog showing the relative values of importance for each feature given to the classifier.  \label{fig:rf_var_import_ext}}
\end{figure*}

Figure \ref{fig:auc_plots_ext} shows the ROC curve for the extended RF.  This is clearly not that of a perfect classifier, but the scale for the False Positive rate is quite small.  Overall, the AUC of the ROC curve is 0.9998, the AUC of the precision recall curve is 0.9898, the F1 number is 0.9817 and the MCC score is 0.9813.  While these numbers are smaller than those from Table \ref{tab:rf_rank} (demonstrating the importance of the IRAC information in this classification problem), they are comparable.  Figure \ref{fig:rf_var_import_ext} shows a feature importance plot for the highest-ranked RF classifier.

\subsection{Extended Candidates \label{sec:ext_cands}}

\begin{figure}
\epsscale{1.2}
\plotone{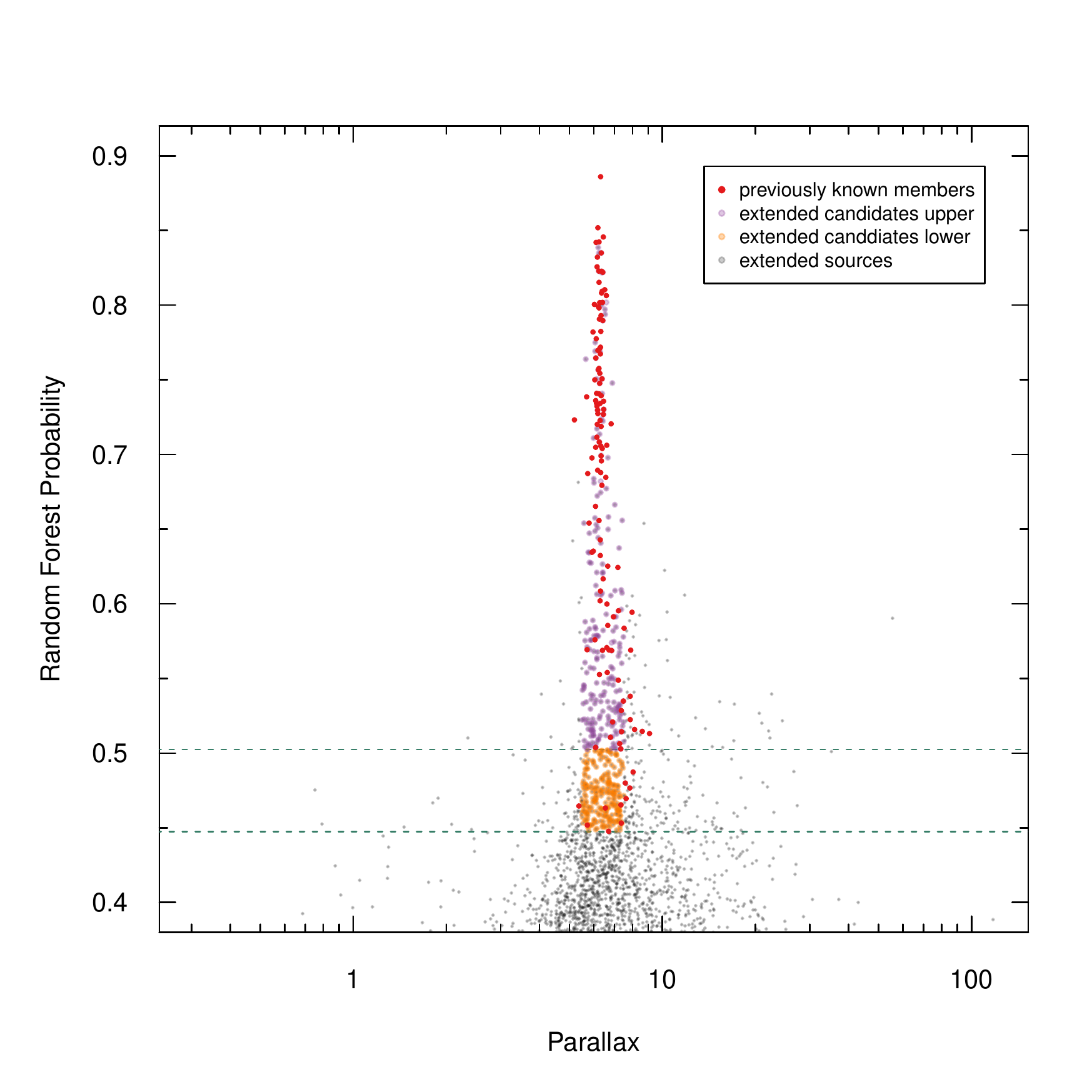}
\caption{Extended RF probability for the extended catalog of sources as a function of the \emph{Gaia} parallax (gray dots).  The previously known members (red dots) are shown along with our proposed new candidate members of the Lupus star-forming region (orange and purple dots).  The green dashed line represents our two probability thresholds for determining a new candidate member at 0.4475 and 0.5025.  \label{fig:rf_prob_ext}}
\end{figure}

We choose two candidate threshold probabilities for the extended catalog.  The first probability threshold (the lower threshold) is 0.4475 to maximize the number of members above the threshold, while minimizing labeled non-member sources.  Figure \ref{fig:rf_prob_ext} does not show member E-21 which has a much lower probability value than the other members.  The lower probability threshold is chosen to be the lower boundary of the members with a greater probability value than member E-21.   The second probability threshold (the upper threshold) is 0.5025.  The confusion matrices for the training sets based on our probability thresholds are given in Tables \ref{tab:conf_matrix_ext1} and \ref{tab:conf_matrix_ext2}.  

Figure \ref{fig:rf_prob_ext} has a wider range of parallaxes among the possible candidates above the chosen probability thresholds than seen in Figure \ref{fig:rf_prob}.  We apply strict parallax limits of greater than 5.5 mas and less than 7.5 mas to be considered an extended candidate in addition to the probability thresholds.  Table \ref{tab:ext_cands_info} lists the catalog information for all of the extended candidates. There are 187 extended candidates above our upper probability threshold (extended candidates upper) and an additional 217 extended candidates that have extended RF probabilities above our lower threshold and below the upper threshold (extended candidates lower).

\begin{table}[h]
\caption{Confusion matrix for the extended RF classifier's training set using a probability threshold of 0.4475. \label{tab:conf_matrix_ext1}}
\noindent
\renewcommand\arraystretch{1.25}
\setlength\tabcolsep{0pt}
\begin{tabular}{l c >{\bfseries}r @{\hspace{0.5em}}c @{\hspace{0.5em}}c}
  \multirow{9}{*}{\rotatebox{90}{\parbox{0.8cm}{\bfseries\centering Predicted}}} & &
    & \multicolumn{2}{c}{\bfseries Actual Label} \\
  & \phantom{'}& & \bfseries m & \bfseries n\\
  & & m$'$ & \MyBox{164} & \MyBox{5} \\[1.5em]
  & & n$'$ & \MyBox{1} & \MyBox{6430}\\
\end{tabular}

\end{table}

\begin{table}[h]
\caption{Confusion matrix for the extended RF classifier's training set using a probability threshold of 0.5025. \label{tab:conf_matrix_ext2}}
\noindent
\renewcommand\arraystretch{1.25}
\setlength\tabcolsep{0pt}
\begin{tabular}{l c >{\bfseries}r @{\hspace{0.5em}}c @{\hspace{0.5em}}c}
  \multirow{9}{*}{\rotatebox{90}{\parbox{0.8cm}{\bfseries\centering Predicted}}} & &
    & \multicolumn{2}{c}{\bfseries Actual Label} \\
  & \phantom{'}& & \bfseries m & \bfseries n\\
  & & m$'$ & \MyBox{154} & \MyBox{2} \\[1.5em]
  & & n$'$ & \MyBox{11} & \MyBox{6433}\\
\end{tabular}

\end{table}

\begin{figure*}
\gridline{
\fig{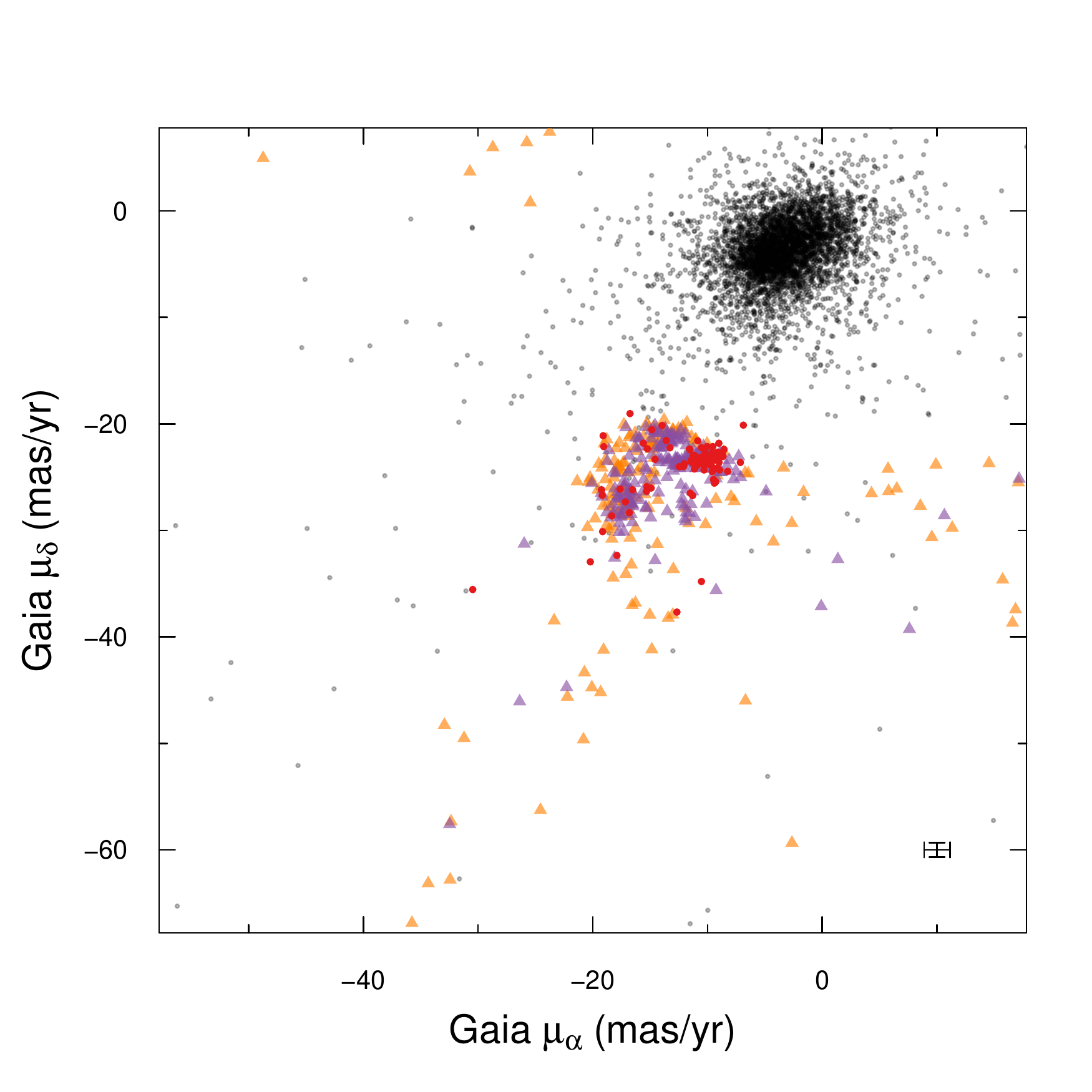}{0.5\textwidth}{(d)}
\fig{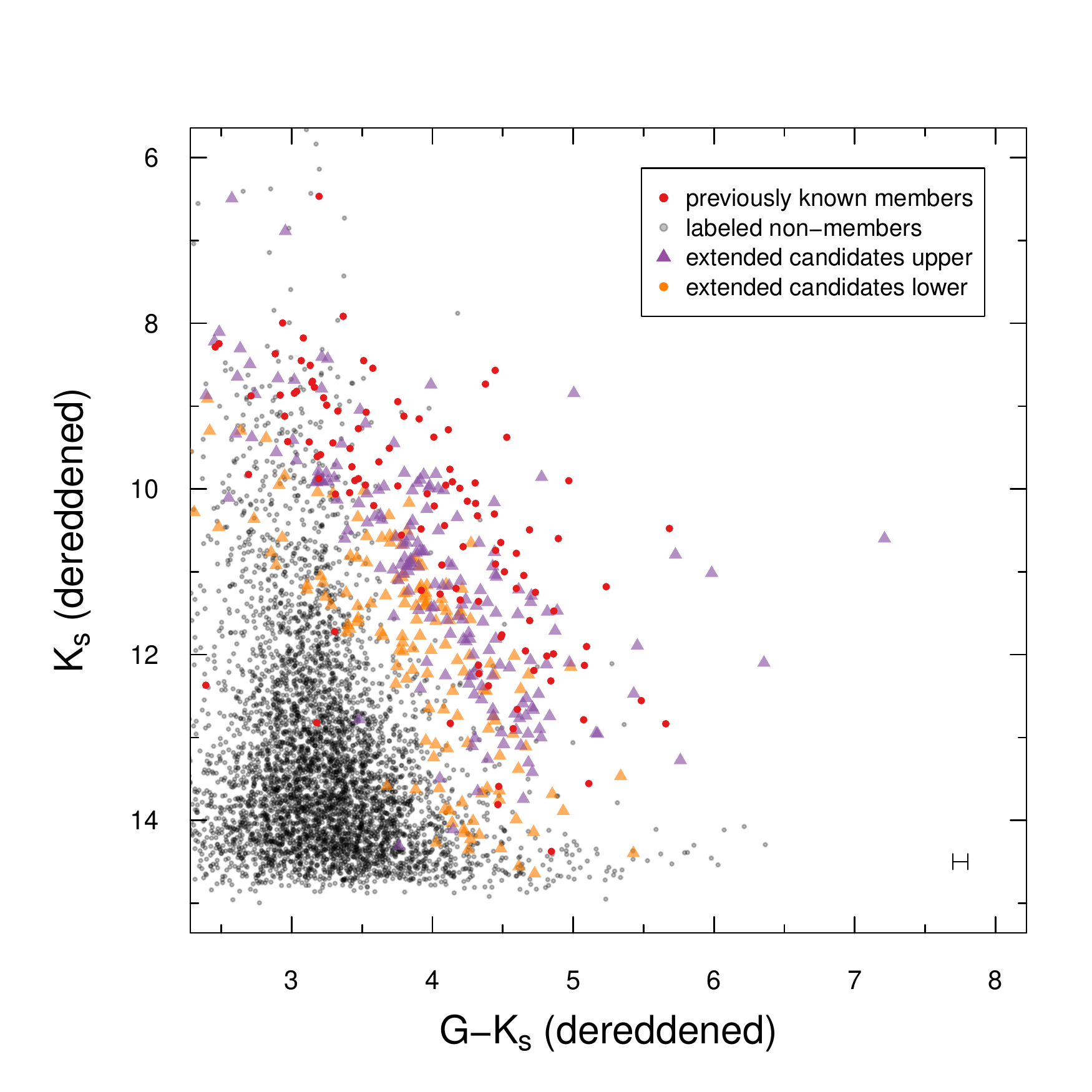}{0.5\textwidth}{(b)}
}
\gridline{
\fig{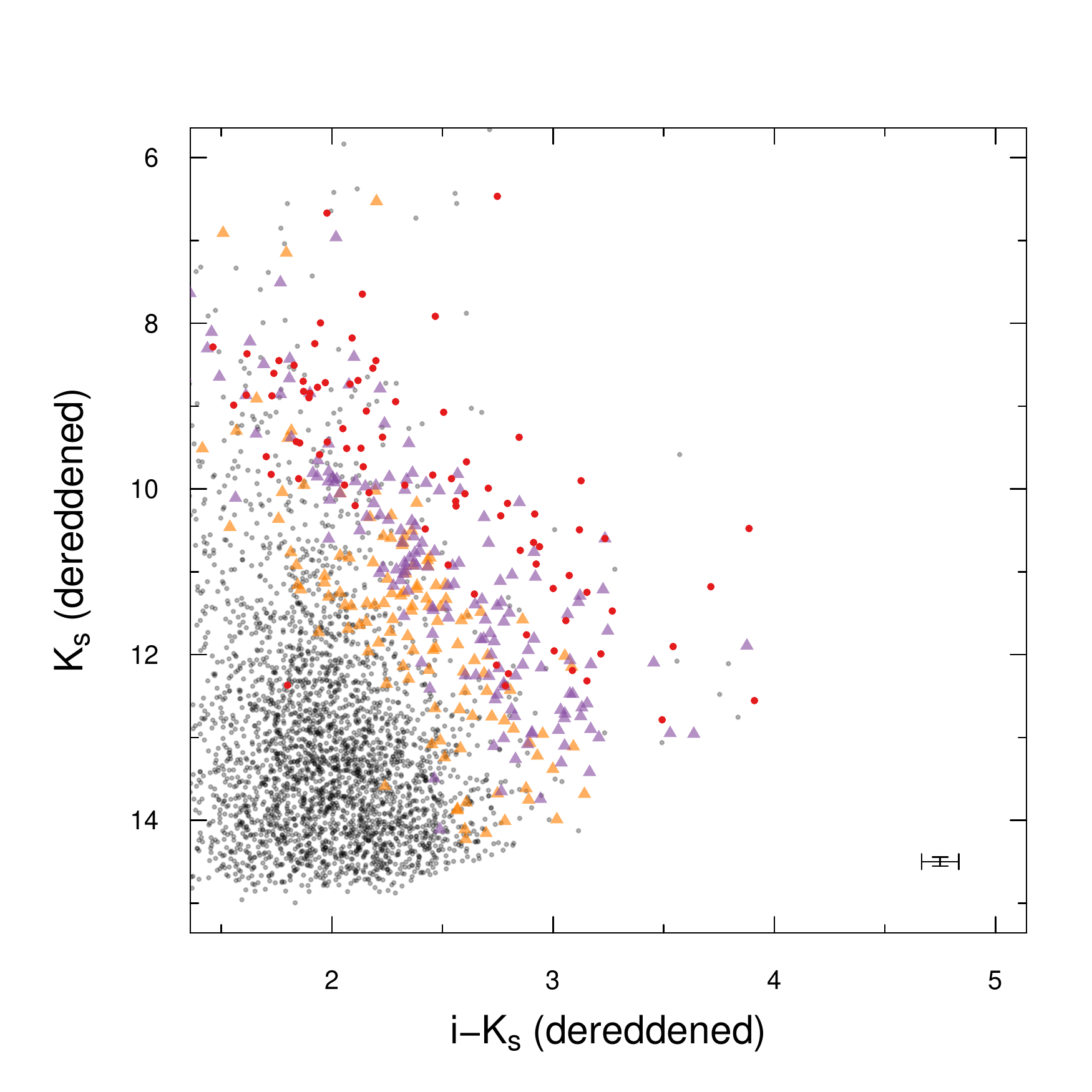}{0.5\textwidth}{(c)}
\fig{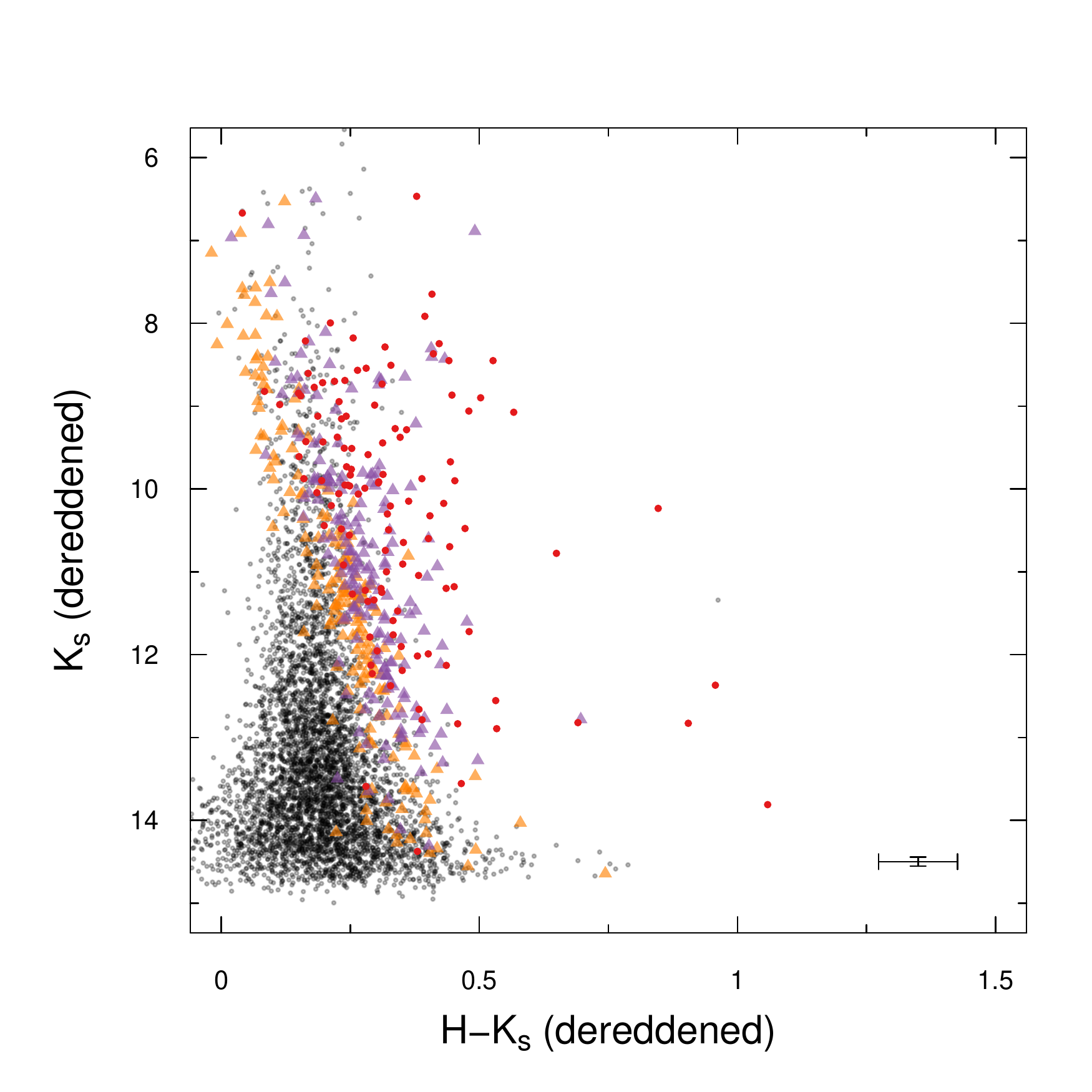}{0.5\textwidth}{(d)}
}
\caption{Proper motion and extinction corrected color-magnitude diagrams for the 2.5\% mix training set (previously identified members as red dots, training set non-members are gray dots) and the proposed candidate members of Lupus (green triangles) picked out by our highest-ranked RF.  The mean error bars for the points are plotted in the bottom right-hand corner of each plot.  \label{fig:rf_ext}}
\end{figure*}

The proper motion plot and the color-magnitude diagrams in Figure \ref{fig:rf_ext} show the distribution of the extended candidate members of the Lupus star-forming region with respect to the previously identified members.  The extended candidates do not cluster as tightly around the previously identified members as the candidates in Section \ref{sec:cands} (compare with Figures \ref{fig:rf_astr} and \ref{fig:rf_cmds}).  There are 106 extended candidates total that have measured proper motion far outside that measured for the previously identified members of the Lupus star-forming region, 16 of which have extended RF probabilities above the upper threshold.  This is significantly more than the five candidates from Section \ref{sec:cands} that had discrepant proper motion.  We do not remove these candidates from our extended candidate catalog because they may represent part of a dynamically ejected population.  Extended candidates with discrepant proper motion are indicated in Table \ref{tab:ext_cands_info} in the 'Label' column by 'pm'.  

\begin{splitdeluxetable*}{lccrrrrBrrrrclll}
\tabletypesize{\scriptsize}
\tablecaption{Catalog information for the candidates picked out by the RF for the extended catalog.   \label{tab:ext_cands_info}}
\tablehead{
\colhead{} & \colhead{RA} & \colhead{DEC} & \colhead{$\mu_{\alpha}$ (\emph{Gaia})} &  \colhead{$\mu_{\delta}$ (\emph{Gaia})}  & \colhead{Parallax} & \colhead{G} & \colhead{\emph{i}} & \colhead{J} & \colhead{H} & \colhead{K} &\colhead{P$_{RF}$\tablenotemark{a}} & \colhead{label\tablenotemark{b}} & \colhead{dis.} & \colhead{SIMBAD} \\
\colhead{} & \colhead{hh:mm:ss.ss} & \colhead{dd:mm:ss.s} & \colhead{mas yr$^{-1}$} & \colhead{mas yr$^{-1}$} & \colhead{mas} & \colhead{mag.}  & \colhead{mag.} & \colhead{mag.} & \colhead{mag.} & \colhead{mag.} &\colhead{} & \colhead{} & \colhead{group\tablenotemark{c}} & \colhead{Name}
}
\startdata
EC-1 & 15:30:13.46 & -32:33:38.9 &  -18.60 $\pm$ 0.21 &  -22.47 $\pm$ 0.16 & 6.85 $\pm$ 0.16 & 14.28 & 12.82 $\pm$ 0.03 & 11.31 $\pm$ 0.02 & 10.69 $\pm$ 0.02 & 10.44 $\pm$ 0.02 & 0.5032 & u &  \nodata   &  \nodata  \\ 
EC-2 & 15:30:25.04 & -32:17:16.7 &  -17.16 $\pm$ 0.16 &  -22.25 $\pm$ 0.11 & 6.71 $\pm$ 0.14 & 14.72 & 13.29 $\pm$ 0.03 & 11.77 $\pm$ 0.02 & 11.10 $\pm$ 0.02 & 10.86 $\pm$ 0.02 & 0.4818 & l & \nodata    & \nodata   \\ 
EC-3 & 15:30:36.58 & -35:26:51.2 &  -17.96 $\pm$ 0.16 &  -22.29 $\pm$ 0.11 & 6.91 $\pm$ 0.11 & 15.47 & 14.44 $\pm$ 0.04 & 12.49 $\pm$ 0.02 & 11.81 $\pm$ 0.02 & 11.58 $\pm$ 0.03 & 0.4786 & l &  \nodata   &\nodata    \\ 
EC-4 & 15:30:36.95 & -32:49:45.3 &  -18.99 $\pm$ 0.30 &  -21.83 $\pm$ 0.22 & 7.15 $\pm$ 0.22 & 17.13 & 15.23 $\pm$ 0.05 & 13.00 $\pm$ 0.03 & 12.37 $\pm$ 0.04 & 12.15 $\pm$ 0.03 & 0.4592 & l & H.K &  \nodata  \\ 
EC-5 & 15:30:54.10 & -32:09:57.0 &   15.72 $\pm$ 0.35 &  -34.63 $\pm$ 0.22 & 6.50 $\pm$ 0.22 & 17.66 & 16.49 $\pm$ 0.08 & 14.53 $\pm$ 0.03 & 13.97 $\pm$ 0.03 & 13.62 $\pm$ 0.04 & 0.4510 & l & pm, i.K &\nodata    \\ 
\enddata
\tablecomments{Two probability thresholds are chosen for this RF, 0.4475 and 0.5025.  Candidates with a RF probability above the upper threshold are marked with a 'u' in the label column.}
\tablenotetext{a}{Probability from the RF classifier for the extended catalog}
\tablenotetext{b}{u if the probability is greater than the upper probability threshold of 0.5025, l if the probability is greater than the lower probability threshold of 0.4475 but less than 0.5025}
\tablenotetext{c}{Discrepant group: pm - inconsistent proper motion with previously identified members, G.K - inconsistent color-index on the $K_{s}$ vs $G$-$K_{s}$ color-magnitude diagram, H.K - inconsistent color-index on the $K_{s}$ vs $H$-$K_{s}$ color-magnitude diagram, i.K - inconsistent color-index on the $K_{s}$ vs $i$-$K_{s}$ color-magnitude diagram}
\tablecomments{Table \ref{tab:members_info} is published in its entirety in the machine-readable format. A portion is shown here for guidance regarding its form and content.}
\end{splitdeluxetable*}

\begin{figure*}
\plotone{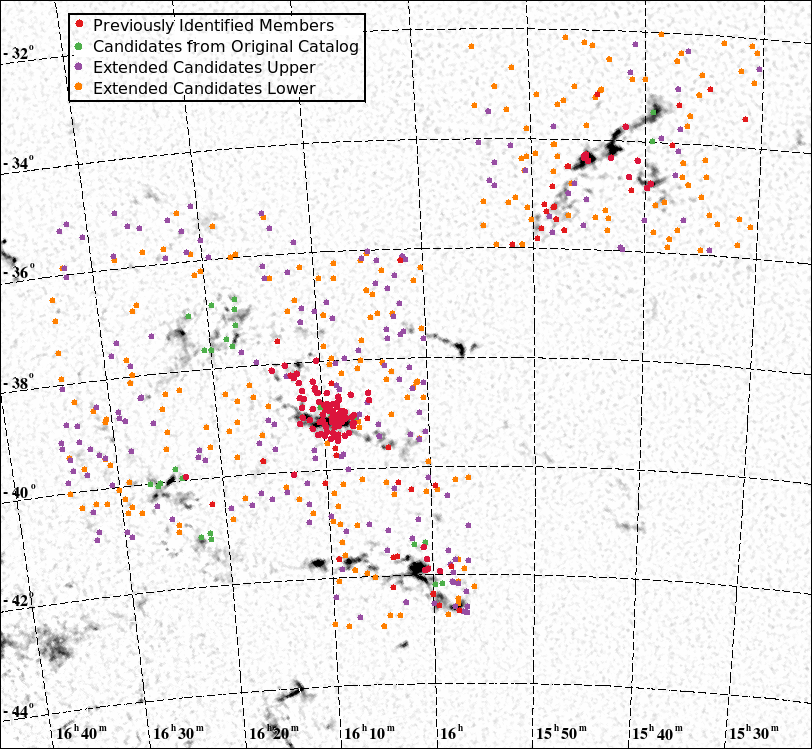}
\caption{Positions of the previously identified members (red), the candidate members from Section \ref{sec:cands} (green), and the extended candidate members from Section \ref{sec:ext_cands} (purple for those above the upper probability threshold, orange for those above the lower probability threshold).  The positions are overlaid on an extinction map of the Lupus star-forming region by \cite{dobashi05} \label{fig:ext_cands_pos}} 
\end{figure*}

The extended candidates positions are also not as well constrained in the color-magnitude diagrams in Figure \ref{fig:rf_ext} to the positions of the previously identified members.  Using the same qualifications to determine a discrepant color-index as in Section \ref{sec:non_mem}, there are 56 extended candidates with inconsistent $G$-$K_{s}$ colors,  79 with inconsistent $i$-$K_{s}$ color indices and 51 with discrepant $H$-$K_{s}$ colors.  There is some overlap between these candidates; the labels for each discrepant group are listed in the 'Label' column in Table \ref{tab:ext_cands_info}.   Figures \ref{fig:rf_ext} (b) and (c) show a gradient in the extended candidates, where the extended candidates closest to the labeled non-members' positions are the extended candidates lower while the extended candidates upper tend to fall much closer to the previously identified members.  The spreading of the extended candidates lower may indicate that these sources belong to an older population of stars while the extended candidates upper may simply represent a young dispersed population.  Figure \ref{fig:ext_cands_pos} shows the positions of the extended candidates with respect to the positions of the Lupus clouds and the previously identified members.  

Our list of extended candidates matches 52 stars with attached references in the SIMBAD database.  These are listed in Table \ref{tab:ext_cands_info}.  Of the 52, three (E-30, E-198 and E-211) have been identified by \citet{comeron09} as candidate members of the Lupus star-forming region.  All three of these previously identified members are above the upper probability threshold for the extended candidate members.  Candidate E-141 was classified as a brown dwarf candidate by \citet{folkes12}.  

Among our extended candidate list, 24 were previously identified to be members of the nearby Scorpius-Centaurus OB association, specifically part of the Upper Centaurus Lupus subgroup of that association (age $\thicksim$ 17 Myr \citealt{mamajek02}).  A list of these candidates and references can be found in Appendix \ref{sec:ext_cands_ucl}. The Lupus star-forming region is very close to the Scorpius-Centaurus association and is believed to be affected by outflows and photoevaporation from the more massive members of Scorpius-Centaurus.  In their paper searching for binary systems, \citet{kraus07} suggest that some members of the Upper Centaurus Lupus subgroup may actually represent an older analog of the Lupus star-forming region.  This lends credence to the idea that the extended RF classifier may be identifying an older population of Lupus.

\section{Discussion \label{sec:discuss}}

\subsection{Spatial and Kinematic Groups in the Lupus Clouds}

Lupus Cloud III in the Lupus star-forming region is one of the richest known concentrations of T-Tauri stars in nearby star-forming regions \citep{comeron08} and also one of the more heavily studied clouds of Lupus.  Most of the previously known members of Lupus lie in Cloud III, with a handful of sources in Clouds I and IV and only one previously identified member in Cloud VI (Figure \ref{fig:pm_final}).  In contrast, of the 27 candidate members picked out by our RF, only four lie in Cloud III, while the majority of the candidates lie in Clouds V and VI. It is not surprising that we did not identify more candidates in Cloud III since it has been so well studied.  Most of the previous searches for members of the Lupus star-forming region were based on accretion signatures \citep{hughes94, krautter97} or infrared excess \citep{comeron03, allen07, merin08, comeron13, muzic14}.  Little to no currently known active star-formation is going on in Clouds V or VI.  The candidates in these clouds may represent a previous generation of star-formation: an older population formed in Clouds V and VI that has lost the obvious signatures of extreme youth.  These candidates in particular are good targets for spectroscopic follow-up to determine membership and set boundaries on their ages.  

The distribution of member and candidate proper motions as measured by \emph{Gaia} are separated into three subgroups (labeled A, B and C in  Figure \ref{fig:rf_astr} (b)).    Cloud III is dominated by kinematic group A, but that is not surprising given that the majority of sources lie in both Cloud III and group A.  Previously identified members in kinematic groups B and C are also present in Cloud III.  Neither Cloud IV or V has any candidates or members from kinematic group B present in them, but this is probably due to the relatively small population of kinematic group B and population of sources within those clouds.  Due to the large amount of spatial mixing among the different kinematic subgroups across such a wide field (approximately 27 by 27 pc for the region shown in Figure \ref{fig:pm_final}), it is possible that the kinematic subgroups are not real and may be due to instrumental error. 

\subsection{Ejected Stars or Contaminants?}

The variability in the astrometric properties of our candidates may be due to the presence of a slightly older, distributed population of pre-main-sequence stars within the Lupus star-forming region \citep{comeron13}.  If star-formation occurs at a continuous rate in a molecular cloud and has been going for a sufficient amount of time, then we would expect to see stars in all stages of evolution from formation to zero-age main-sequence around molecular clouds.  However, while there are large populations of very young stars ($\leq$ a few Myr) associated with star-forming regions, we do not generally see corresponding populations of older pre-main-sequence stars \citep{feigelson96}.  

In the Taurus-Auriga star-forming region (similar age and distance as Lupus) for example, there is some discussion about the completeness of the census due to a possible dispersed population of older pre-main-sequence stars.  \citet{kraus17} found a large dispersive population, including 81  sources that had been omitted from previous censuses.  
Most of those sources are rejected as members of the Taurus-Auriga star-forming region by \citet{luhman18}, who does not find evidence for a dispersive population in Taurus.  
This discrepancy can be largely traced to the definitions used to label members of the star-forming region.  
\citet{kraus17} weighted youth signatures in their sources more heavily than kinematic constraints while \citet{luhman18} relied upon strict kinematic consistency in creating their membership list.  

The Lupus star-forming region lies above the plane of the Galactic disk stretching from $b \approx 6^\circ$ to $b \approx 17^\circ$. Especially at the lower Galactic latitudes, contamination of the candidates by background stars is a possible concern.  Asymptotic giant branch (AGB) stars are the most likely contaminants because their cool colors can mimic those of reddened young stellar objects.  To estimate the amount of contamination in our candidate sample, we use our RF and our candidate probability threshold to classify simulated data from the \citet{robin03} Galaxy model\footnote{\label{footnote_galaxy_model}http://model.obs-besancon.fr/}.  Given the photometric limits of our on-cloud catalog, the classifier labels no AGB sources as candidate members of Lupus.  While Galactic count models are not as accurate for low-mass sources \citep{spezzi11} this does nevertheless demonstrate that AGB stars are not contributing heavy contamination to our on-cloud candidates.

We use the \citet{robin03} Galaxy model\textsuperscript{\ref{footnote_galaxy_model}} to estimate the contamination of extended candidates by background AGB stars.  The simulated data labeled as candidates by our extended RF classifier are contaminated between 2.05\% and 2.75\% by AGB stars.  The extended RF has greater difficulty than our original classifier in separating members and non-members due to the lack of IRAC photometry.  However, even though the contamination levels by background AGB stars are significantly higher for our extended RF, the levels are still low enough that we can positively state that AGB stars are not contributing heavy contamination to our extended candidates. Extended candidate E-147 was identified by \citet{ruizDern18} as a member of the red giant clump based on photometric and parallactic criteria.  It also has a measured spectral type of K1/2 III \citep{houk86}.  Without specific gravity measurements and measurements of youth indicators (Li abundance, H$\alpha$ emission, etc.) it is difficult to separate young low-mass stars from giants since they occupy the same space in the HR diagram, though our RF classifiers do a fairly good job.

As discussed in Section \ref{sec:cands}, the large disparity between the number of sources with large parallaxes and those with small parallaxes used to train the RF can make sources with large parallaxes seem to be more reasonable possible candidates.  
The distribution of characteristics of the intrinsic population of the catalog covering the five cloud regions may skew the probability assignment of the RF.  
Figures \ref{fig:rf_astr}  (a) and (b) both show that candidates 1, 6, 9, 25 and 26 all have a larger magnitude of proper motion (measured by both IRAC and \emph{Gaia}) than the members and other candidates.  However, it is also evident that concentration of labeled non-members is not large near these high proper motion candidates, especially compared to the concentration of labeled non-members around (0, 0) mas yr$^{-1}$. 
The non-uniformity of the distribution of the characteristics of the labeled non-member sources passed to our RF could be remedied by injecting a population of synthetic non-members into the on-cloud catalog, but such a population would not accurately reflect the underlying structure of the true data. 
Furthermore, the synthetic population of non-members may mask true members, such as members of a distribution of older pre-main-sequence stars that have drifted from the initial molecular cloud or possibly ejected from the cloud shortly after formation.  

\subsection{On-Cloud Candidates are Mostly Disk Free \label{sec:cand_disk}}

The presence of a disk can be inferred through the presence of infrared excess in a source's spectral energy distribution (SED) or by its Lada class  (defined by \citealt{lada84}, extended by \citealt{greene94}).  We use the previous disk determinations in \citet{merin08} to guide our identification of disk presence around our on-cloud candidates. Overall, \citet{merin08} found a disk fraction of 70\%-80\% for the Lupus star-forming region.  Table \ref{tab:members} includes the Lada class of the previously identified sources in Lupus when available in the literature.

\citet{merin08} presents the c2d IRAC observations and analysis for the Lupus star-forming region.  Their member list (see \citet{merin08} Table 5) is 159 sources long; 78 of their members overlap with our previously identified membership list.  Some of the member discrepancy lies in coverage; they include eight off-cloud IRAC observing runs not included in our study.  \citet{merin08} has a lower threshold for classifying an object as a member, requiring only that the SED be indicative of youth, while we require that members be spectroscopically confirmed.  

\begin{figure*}
\plotone{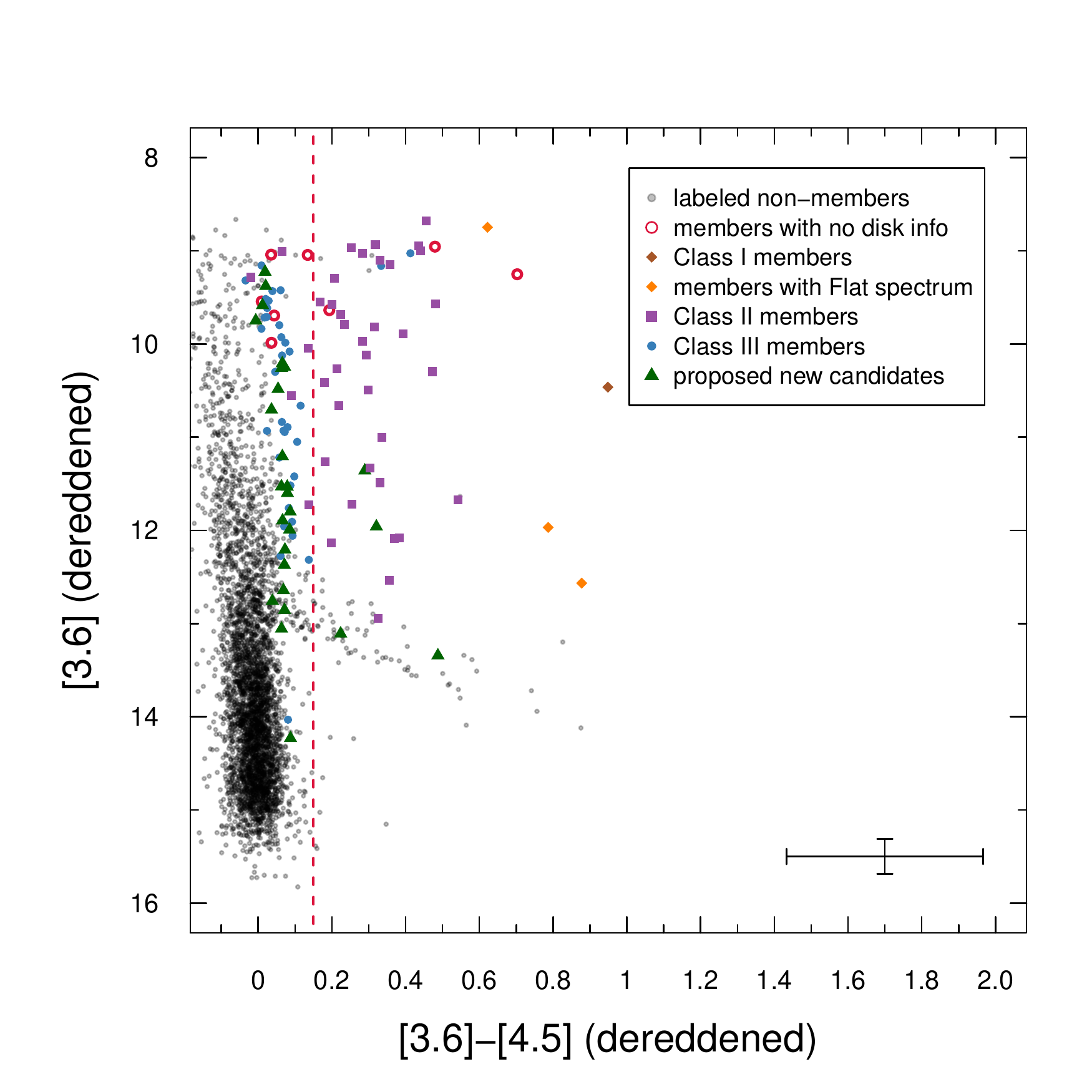}
\caption{Color-magnitude diagram (from Figure \ref{fig:rf_cmds} (d)) with the previously identified members of Lupus formatted to reflect their Lada class (see Table \ref{tab:members}).  The vertical line at 1.5 represents the break where we take candidates to the right of the line to have Class II disks and candidates to the left not have an accretion disk.    \label{fig:disk_frac}}
\end{figure*}

Figure \ref{fig:disk_frac} demonstrates this through the separation of embedded sources (Class I and flat sources) and sources with disks (Class II sources) from those sources with either a debris disk or no disk (Class III sources) as a function of infrared color.  The vertical line separates (as best as possible) the sources with infrared excess (to the right of the line) from those sources with no infrared excess (to the left). Only two of the previously identified Class III sources lie to the right: Members 49 and 64 (see Table \ref{tab:members}).  The separation line allows the classification of a disk presence around eight of our previously identified members without any disk indication in the literature (see Table \ref{tab:members}) and all 27 of our candidates (see Table \ref{tab:cands_photometry}).  

To calculate disk fraction for the Lupus star-forming region, we define any source with a Lada class less than or equal to Class II or infrared excess as disk-bearing and Class III sources as disk-free.  Previously identified members for which no information regarding infrared excess is available are not included in this analysis.  Sixty-two of the previously identified members are disk-bearing compared to 48 disk-free members, giving a disk fraction of 56\% which is lower than that found by \citet{merin08}.  Including our candidate members (four of the 27 show indication of a disk from Figure \ref{fig:disk_frac}) lowers the disk fraction to 48\%.  The high fraction of Class III objects agrees with the findings of \citet{spezzi11}, who applied the same procedures to finding Lupus candidates and measuring disk fractions as the c2d team, but they focused their study on Lupus Clouds V and VI rather than I, III, and IV like \citet{merin08}.  \citet{spezzi11} also deduced that the stellar populations of Lupus V and VI represents a slightly (few Myr) older population of star formation.

Our membership list is based on spectroscopically confirmed members of the Lupus star-forming region rather than SEDs and photometric indications.  Therefore, our member list is already more likely to contain older stars and disk-free stars than the membership list used by \citet{merin08}.  Our candidate list consists mainly of sources toward Lupus Clouds V and VI where no current star-formation is known to be occurring.  If these clouds do indeed represent a population of slightly older stars than the populations of Lupus Clouds I, III and IV, then it would be expected that most of the disks around these stars would have dissipated.  One caveat to consider when classifying disk presence for our sources is the large error compared with the scale of the color indices in Figure \ref{fig:disk_frac} which could cause sources to be misclassified.  

\begin{figure}
\epsscale{0.75}
\plotone{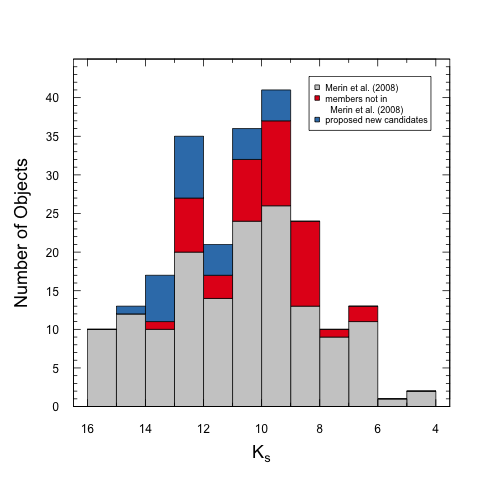}
\caption{Distribution of sources found by the c2d legacy survey \citep{merin08} (gray), the previously identified members in Table \ref{tab:members} not in \citet{merin08} (red) and our proposed candidate members of Lupus (blue, Table \ref{tab:cands_astrometry}). \label{fig:hists}}
\end{figure}

Figure \ref{fig:hists} compares the distribution of sources labeled as members by \citet{merin08} (based on SED alone) with the updated distribution of Lupus sources including \citet{merin08} members, the previously identified members used in this study (Section \ref{sec:mem}) and our proposed candidate members.  The peaks in the updated histogram occur in the same locations, but there is a much more dramatic difference between the number of faint sources and the peak in the updated histogram (Figure \ref{fig:hists}).  Our addition to the \citet{merin08} catalog adds no bright sources, furthering the widely known overabundance of faint sources in the Lupus star-forming region (\citealt{merin08}; \citealt{comeron08} and references therein).  

\section{Conclusions}

We combine measured IRAC proper motion, \emph{Gaia} proper motion and parallax and photometry from IRAC, \emph{Gaia}, 2MASS and DENIS for stars in the Lupus star-forming region observed with IRAC.  We compare multiple RF classifiers to find the best RF tuning parameters as well as the best input data qualities to separate known members from labeled non-members of the Lupus star-forming region.  Finally, using our best performing RF, we present a list of candidate members of the Lupus star-forming region for spectroscopic follow-up in the future.  Our candidates generally have properties that are consistent with those of members of the Lupus star-forming region.

The majority of our candidates lie in Lupus Clouds V and VI.  There is only one previously identified candidate in Lupus Cloud V.  These sources probably represent a slightly older population than that present in the more active clouds (Clouds I, III and IV).  The estimated disk fraction of our candidate sources is also low, lending credence to the older population postulation. 

We find that Lupus III has a mean distance of 156.5 pc (closer than the previous distance measurements of 200 pc: \citealt{comeron08}).  The distances to sources associated with Lupus III range from 110.1 pc to 192.6 pc which while large, do agree with the previously measured thickness of the cloud of 80 pc, though this measurement has a large uncertainty \citep{comeron08}.   The distance to the other Lupus Clouds (mean of 148.8 pc) agrees well with previous measurements of 150 pc. The members of Lupus cover a very similar range of distance.   The distance gap between the clouds which has been used in the previous literature is not present in the members of the Lupus star-forming region.  Overall, the majority of the sources in Lupus have a distance between 152.6 and 162.7 pc.  

When training a RF, we find that the form of the feature inputs is very important in building the most effective classifier.  All of our top classifiers have the proper motion data and the photometry data in the form that is most physically relevant when grouping and classifying sources.  
The RFs that use the full member list, rather than a constrained list, are better at modeling the variety of properties seen in the membership class.  We find that class imbalance compensation is also important for RF classification, especially when the member class is only 2.5\% of the entire training set.  Overall, the balanced RF method performs slightly better for class imbalance compensation.  
We apply the same RF tuning scheme to a RF built for an extended catalog of sources.  The extended RF has similar performance metrics to the original RF, which demonstrates that the training qualities extend to classifiers for related problems using different source catalogs.  

We present a catalog of extended candidate members of the Lupus star-forming region with two probability thresholds chosen for the extended RF.  The properties of the extended candidates are much more dispersed than those in our original candidate list, indicating that these candidates may not be as good for spectroscopic follow-up.  Twenty-four of our 404 extended candidates have been previously identified to be members of the nearby subgroup of the Scorpius-Centaurus OB association, Upper Centaurus Lupus.  \citet{kraus07} suggest that Upper Centaurus Lupus is actually an older population of stars dispersed from Lupus, indicating that our extended RF classifier identified some dispersed older stars of the Lupus star-forming region.

\acknowledgements

We thank E. Feigelson for his guidance on statistical methodology and applications and valuable discussions.
We thank K. Luhman and T. Esplin for their guidance in executing the ``IRACpm" package.   
This work is based in part on observations made with the \textit{Spitzer Space Telescope} and the IPAC Infrared Science Archive (IRSA), which are operated by the Jet Propulsion Laboratory, California Institute of Technology and Caltech under contract with NASA.  
This publication makes use of data products from the Two Micron All Sky Survey, which is a joint project of the University of Massachusetts and the Infrared Processing and Analysis Center/California Institute of Technology, funded by the National Aeronautics and Space Administration and the National Science Foundation. 
 The DENIS project has been partly funded by the SCIENCE and the HCM plans of the European Commission under grants CT920791 and CT940627. It is supported by INSU, MEN and CNRS in France, by the State of Baden-W\"urttemberg in Germany, by DGICYT in Spain, by CNR in Italy, by FFwFBWF in Austria, by FAPESP in Brazil, by OTKA grants F-4239 and F-013990 in Hungary and by the ESO C\&EE grant A-04-046.  Jean Claude Renault from IAP was the project manager.  Observations were carried out thanks to the contribution of numerous students and young scientists from all involved institutes, under the supervision of  P. Fouqu\'e, survey astronomer resident in Chile.  
This work has made use of data from the European Space Agency (ESA) mission {\it Gaia} (\url{https://www.cosmos.esa.int/gaia}), processed by the {\it Gaia} Data Processing and Analysis Consortium (DPAC, \url{https://www.cosmos.esa.int/web/gaia/dpac/consortium}). Funding for the DPAC has been provided by national institutions, in particular the institutions participating in the {\it Gaia} Multilateral Agreement.

\appendix

\section{Distance to Lupus Clouds \label{sec:dist_clouds}}

\begin{figure*}
\gridline{\fig{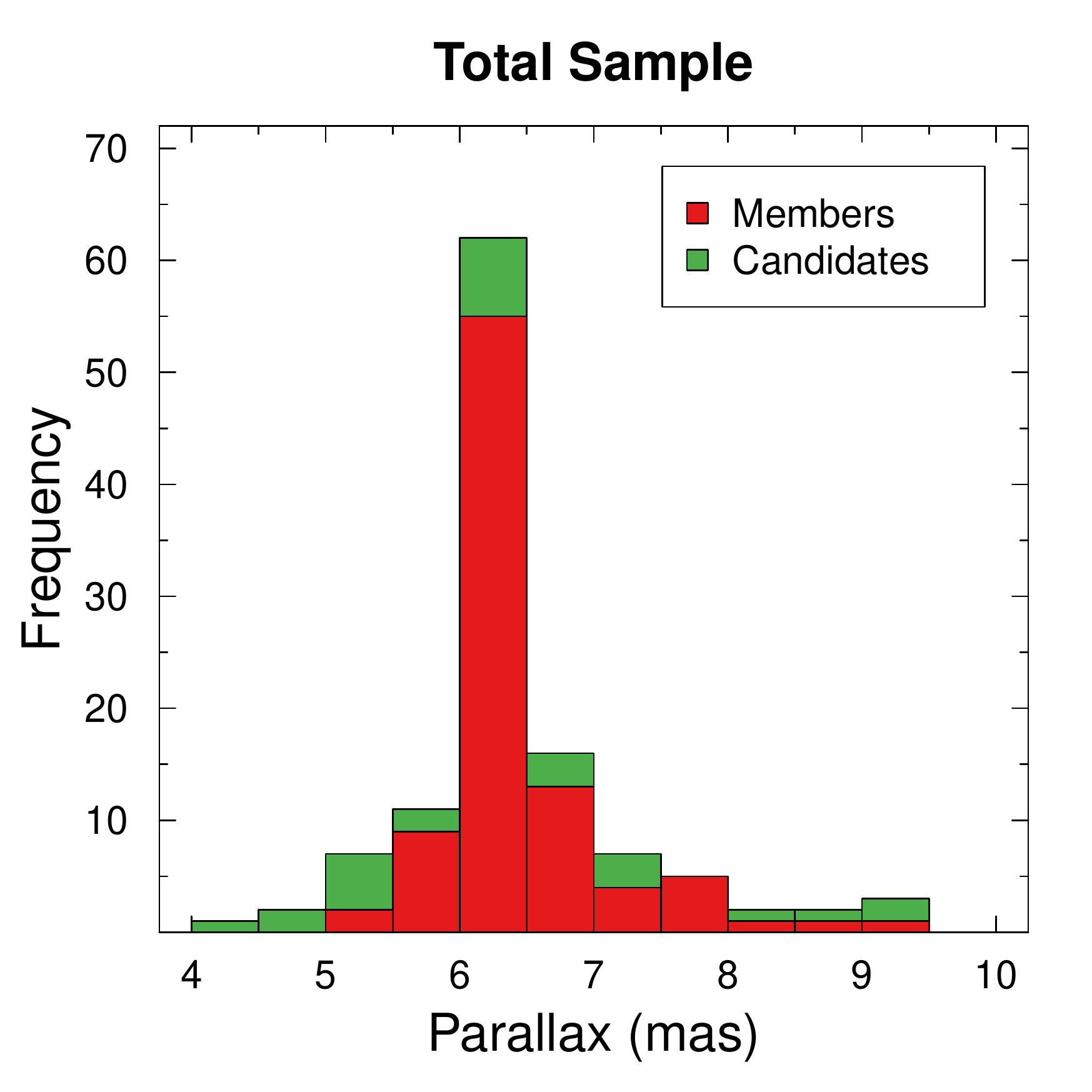}{0.3\textwidth}{(a)}
\fig{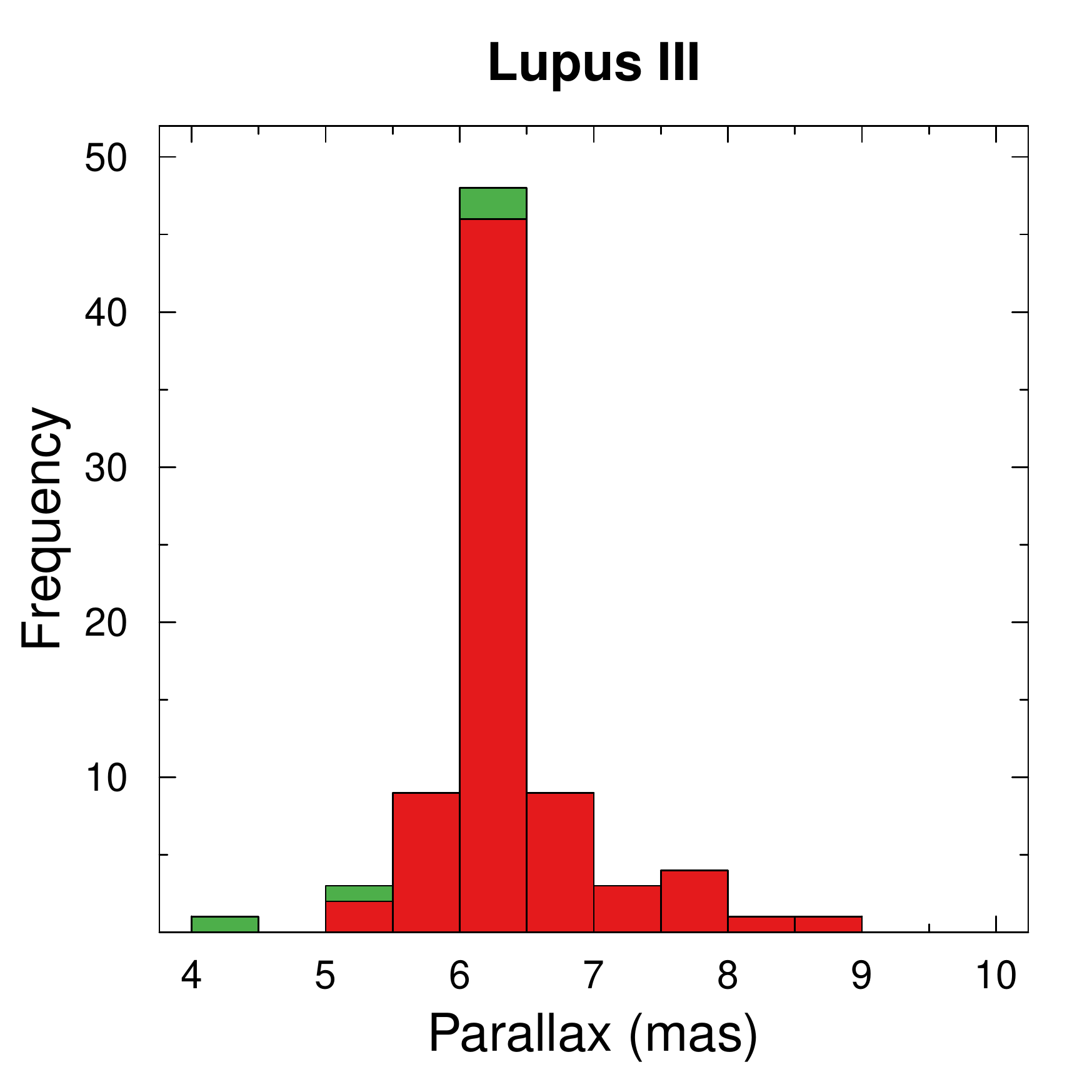}{0.3\textwidth}{(b)}
\fig{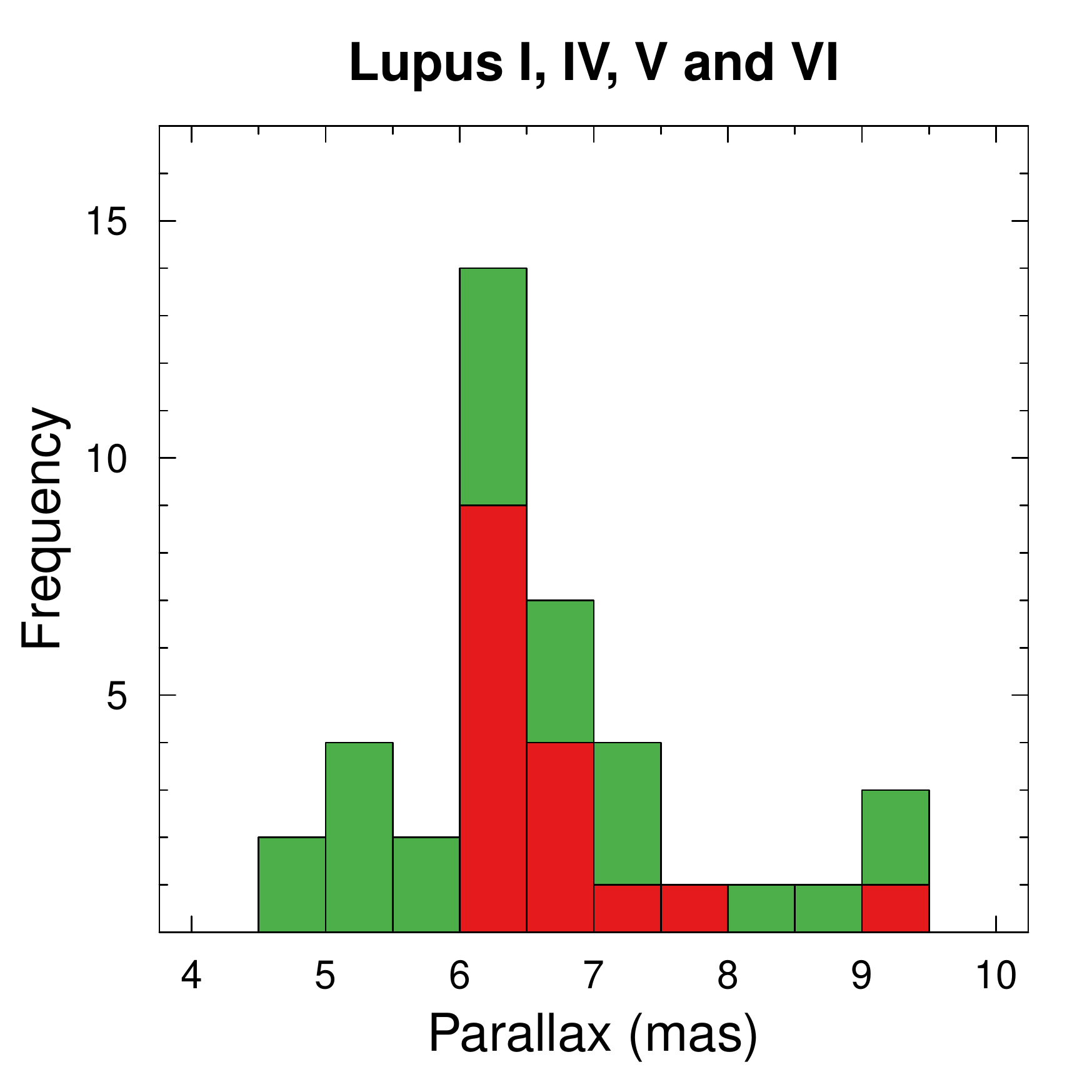}{0.3\textwidth}{(c)}
}
\caption{Distribution of the \emph{Gaia} parallax measurements for the previously identified members (red) and the proposed on-cloud candidate members of Lupus  (green) identified by our highest-ranked RF classifier. The entire distribution of sources for all of the clouds in the Lupus star-forming region are shown in (a), while the distribution for Cloud III (b) has been separated from the other members and candidates in (c).     \label{fig:par_hists}}
\end{figure*}

The standard practice in the literature is to take the distance to Lupus III as 200 pc and the distance to all of the other clouds in Lupus as 150 pc.  The distribution of the \emph{Gaia} parallaxes for the members of the Lupus star-forming region and our candidate members (Figure \ref{fig:par_hists} (a)) shows a very sharp peak in the 6 - 6.5 mas bin (153.8 - 166.7 pc), with a large overall spread between 9.5 and 4 mas.  The parallax distribution just for Cloud III (Figure \ref{fig:par_hists} (b)) shows the same peak.  This is not surprising since most of the  members of Lupus are in Cloud III, but it does show that using a distance of 200 pc is incorrect.  There are only, at most, three sources with a distance of 200 pc and only one candidate member that has a distance farther than 200 pc.  The distribution of parallaxes for the other Lupus Clouds (Figure \ref{fig:par_hists} (c)) have the same peak in parallax distribution as Lupus III indicating that the entire Lupus star-forming region occupies the same range of parallaxes.  The depth of the Lupus star-forming region has been estimated before to be $\thicksim$80 pc \citep{makarov07}.  From the previously identified members, the depth of Lupus is 82.5 pc from the nearest previously identified member in the catalog to the farthest.  Including our candidate members extends that depth to 120 pc.  The range of distances to the other Lupus Clouds is pushed back by our proposed candidate members (candidates compose the eight farthest members in Figure \ref{fig:par_hists} (c)).  Given the distribution of \emph{Gaia} parallax measurements of the previously identified members, the median distance of 155 pc is appropriate for the entire Lupus star-forming region.

\section{Notes on Named Candidate Sources\label{sec:named_cands}}

\subsection{2MASS J15590456-4210583 and 2MASS J16065870-3904051}

2MASS J15590456-4210583 and 2MASS J16065870-3904051 were both identified as candidate members of Lupus by \citet{comeron09}.  
\citet{comeron09} performed a wide field photometry survey of Lupus and identified candidate members by fitting model spectra to the SED of their sources. 

\subsection{2MASS J16171811-3646306, 2MASS J16172485-3657405 and 2MASS J16192684-3651235}

2MASS J16171811-3646306, 2MASS J16172485-3657405 and 2MASS J16192684-3651235 were identified as young stellar objects by \citet{spezzi11} through model fitting to their SEDs. 
 \citet{spezzi11} created the SEDs by combining photometry from IRAC ([3.6], [4.5], [5.8], [8.0]), MIPS (24 $\mu$m and 80 $\mu$m), NOMAD (B, V and R) along with DENIS ($i$, $J$ and $K_s$) and 2MASS ($J$, $H$ and $K_s$).  
Based on the slopes of the SEDs, they classified 2MASS J16171811-3646306 and 2MASS J16192684-3651235 as class II objects (young stellar objects embedded within an accretion disk) and 2MASS J16172485-3657405 as a class III object (young stellar object with a debris disk).  
These same classifications were also determined independently by \citet{dunham15} by fitting models to SEDs for 2MASS J16171811-3646306, 2MASS J16172485-3657405  and 2MASS J16192684-3651235 as part of the \emph{Spitzer Space Telescope} legacy surveys ``cores to disks" and ``Gould Belt".  

Recently, 2MASS J16171811-3646306, 2MASS J16172485-3657405 and 2MASS J16192684-3651235 have been examined by \citet{manara18} who used data from the second data release of \emph{Gaia} to examine candidate members of the Lupus star-forming region.  
\citet{manara18} labels all three objects as likely members of the Lupus star-forming region based on their parallax, proper motion and photometry. 

\subsection{2MASS J16081739-3901062}

2MASS J16081739-3901062 was detected as an x-ray source by \citet{gondoin06}, but no follow-up analysis was performed at the time.  
\citet{comeron09} identified 2MASS J16081739-3901062 as a candidate member of the Lupus star-forming region through SED fitting as part of a wide-field photometric survey.  
\citet{lopezMarti11} did a proper motion survey of the Lupus star-forming region using virtual observatory tools and labeled 2MASS J16081739-3901062 as a probable member of Lupus based on proper motion.  

\subsection{2MASS J16090849-3902133}

2MASS J16090849-3902133 was identified as a candidate member of Lupus by two different photometric surveys: \citet{lopezMarti05} and \citet{comeron09}.  
Both surveys combined photometric data from well known surveys such as 2MASS and DENIS with pointed optical observations to create SEDs for model fitting and analysis.  \citet{comeron11} also labeled 2MASS J16090849-3902133 as a candidate member of the Lupus star-forming region based on deep pointed photometric observations. 

\section{Extended Candidates in Upper Centaurus Lupus\label{sec:ext_cands_ucl}}

E-183 is listed as a member of the Scorpius-Centaurus OB association by \citet{deGeus90}.   \citet{hoogerwerf00} compiled a membership list of the Scorpius-Centaurus OB association.  Seventeen of the extended candidates are listed as members of the Upper Centaurus Lupus subgroup: E-168, E-202, E-208, E-209, E-210, E-220 (see also \citealt{chen12}), E-242, E-247, E-282, E-303, E-306, E-314, E-315, E-346, E-349, E-376, and E-397.  E-65 is identified as a member of Scorpius-Centaurus by \citet{kohler00}.  In addition to E-220 (previously identified), \citet{chen12} also lists E-190 as a member of the Upper Centaurus Lupus subgroup.  \citet{song12} spectroscopically confirms E-40 as a member of the Upper Centaurus Lupus subgroup.  E-119, E-216 and E-349 (previously listed by \citealt{hoogerwerf00}) are all included in \citet{pecaut16}.  Finally, in addition to confirming E-397 (previously listed by \citealt{hoogerwerf00}), \citet{roser18} identifies E-400 as a member of Upper Centaurus Lupus.

\end{document}